\documentclass[pra,%
 reprint,%
 twocolumn,
superscriptaddress,
nofootinbib,
longbibliography]{revtex4-2}

\usepackage{array,amsmath,amsfonts,amssymb,dsfont,tabularx,multirow,blkarray,bigstrut}

\usepackage{hhline}
\usepackage[T1]{fontenc} \usepackage{ae} \usepackage{color}
\usepackage{bbold}
\usepackage{enumitem}
\usepackage{lipsum}
\usepackage{url}
\usepackage[colorlinks=true,linkcolor=blue,citecolor=blue,urlcolor=blue,plainpages=false,pdfpagelabels]{hyperref}

\usepackage{color}
\usepackage[dvipsnames]{xcolor}
\usepackage{mathtools}
\usepackage{bbm}

\newcolumntype{C}[1]{>{\centering\arraybackslash$}p{#1}<{$}}

\usepackage{graphicx}
\usepackage{soul}
\usepackage{balance}

\usepackage{amssymb}
\usepackage{braket}
\usepackage{pifont}

\newcommand{\appropto}{\mathrel{\vcenter{
  \offinterlineskip\halign{\hfil$##$\cr
    \propto\cr\noalign{\kern2pt}\sim\cr\noalign{\kern-2pt}}}}}

\usepackage{makecell,tabularx}

 \def\1{\mathchoice{\rm 1\mskip-4.2mu l}{\rm 1\mskip-4.2mu l}{\rm
1\mskip-4.6mu l}{\rm 1\mskip-5.2mu l}}

 \newcommand{\id}

\usepackage{gensymb}

\begin{document}

\title{Correcting heading errors in optically pumped magnetometers through microwave interrogation}

\author{C.~Kiehl}\email{christopher.kiehl@colorado.edu}
\affiliation{JILA, National Institute of Standards and Technology and University of Colorado, Boulder, Colorado 80309, USA}
\affiliation{Department of Physics, University of Colorado, Boulder, Colorado 80309, USA}
\author{T.~S.~Menon}
\affiliation{JILA, National Institute of Standards and Technology and University of Colorado, Boulder, Colorado 80309, USA}
\affiliation{Department of Physics, University of Colorado, Boulder, Colorado 80309, USA}
\author{D.~P.~Hewatt}
\affiliation{JILA, National Institute of Standards and Technology and University of Colorado, Boulder, Colorado 80309, USA}
\affiliation{Department of Physics, University of Colorado, Boulder, Colorado 80309, USA}
\affiliation{Paul M. Rady Department of Mechanical Engineering, University of Colorado, Boulder, Colorado 80309, USA}
\author{S.~Knappe} 
\affiliation{Paul M. Rady Department of Mechanical Engineering, University of Colorado, Boulder, Colorado 80309, USA}
\affiliation{FieldLine Medical, Boulder CO 80301, USA}
\affiliation{FieldLine Industries, Boulder CO 80301, USA}
\author{T.~Thiele} 
\affiliation{Zurich Instruments AG, CH-8005 Zurich, Switzerland}
\author{C.~A.~Regal} 
\affiliation{JILA, National Institute of Standards and Technology and University of Colorado, Boulder, Colorado 80309, USA}
\affiliation{Department of Physics, University of Colorado, Boulder, Colorado 80309, USA}

\begin{abstract}
We demonstrate how to measure \textit{in situ} for heading errors of optically pumped magnetometers (OPMs) in the challenging parameter regime of compact vapor cells with imperfect optical pumping and high buffer gas pressure. For this, we utilize microwave-driven Ramsey and Rabi frequency spectroscopy (FS) to independently characterize scalar heading errors in free induction decay (FID) signals. Both of these approaches suppress 5-nT inaccuracies in geomagnetic fields caused by nonlinear Zeeman (NLZ) shifts in FID measurements to below 0.6 nT. For Ramsey FS, we implement short periods of microwave interrogation within a \(\pi/2-t_R-3\pi/2\) Ramsey interferometry sequence, effectively circumventing systematic errors from off-resonant driving. Conversely, Rabi FS leverages an atom-microwave Hamiltonian for accurate modeling of Rabi oscillation frequencies, achieving a measurement precision down to 80 pT$/ \sqrt{\text{Hz}}$ that is limited primarily by technical microwave noise. We show that the fundamental sensitivity of Rabi FS is 30 pT/$\sqrt{\text{Hz}}$ with our vapor cell parameters through a Cramér-Rao lower bound (CRLB) analysis. This work paves the way for future investigations into the accuracy of hyperfine structure (HFS) magnetometry and contributes to the broader applicability of OPMs in fields ranging from navigation and geophysics to space exploration and unexploded ordinance detection, where heading error mitigation is essential.
\end{abstract}
    
\maketitle
\raggedbottom
\section{introduction}
Optically pumped magnetometers (OPMs) are state-of-the-art sensors that can reach sensitivities below 1 fT $/\sqrt{\text{Hz}}$~\cite{kominis2003subfemtotesla,budker2007optical,dang2010ultrahigh,sheng2013subfemtotesla}, enable precise detection of biomagnetic signals~\cite{bison2003laser,xia2006magnetoencephalography,broser2018optically}, and push the boundaries for scientific exploration by aiding in searches for permanent electron dipole moments~\cite{pendlebury1984search,ayres2021design} and dark matter~\cite{pospelov2013detecting,afach2021search}. Practical use of OPMs in geomagnetic fields such as navigation~\cite{psiaki1993ground,canciani2016absolute}, geophysics~\cite{friis2006swarm,stolle2021special}, space~\cite{dougherty2004cassini,korth2016miniature,bennett2021precision}, and unexploded ordinance detection~\cite{billings2004discrimination,prouty2016real} requires addressing systematic errors that depend on the orientation of the sensor with respect to the magnetic field known as heading errors. For the most common OPMs made of alkali atoms the dominant heading error at geomagnetic fields is of the order of 10 nT. This systematic error manifests from unknown strengths of unresolved frequency components in the magnetometer signal arising from nonlinear Zeeman (NLZ) shifts from each of the ground-state hyperfine manifolds~\cite{alexandrov2003recent,lee2021heading}.

Only in regimes of narrow magnetic resonances~\cite{acosta2006nonlinear} and high spin polarization can this heading error be accurately modeled to 0.1 nT~\cite{lee2021heading}. In vapor cells using microelectromechanical systems (MEMS) technology, a compact vapor cell solution that is well suited for scalable mass production, these regimes often become unfeasible. This is due to line broadening from atomic collisions and the challenges associated with achieving fast, high-fidelity optical pumping with modest pump powers. Various other approaches have been developed to mitigate heading error including spin locking~\cite{bao2018suppression,bao2022all}, light polarization modulation~\cite{oelsner2019sources}, double-pass configurations~\cite{rosenzweig2023heading}, double-modulated synchronous pumping~\cite{seltzer2007synchronous}, and leverage of tensor light shifts~\cite{jensen2009cancellation}, but all these approaches neglect frequency shifts arising from the different Zeeman resonances between the $F=I\pm 1/2$ manifolds and have their own practical challenges. Furthermore, methods that utilize higher-order polarization moments~\cite{zhang2023heading,acosta2008production,yashchuk2003selective} are not feasible with high buffer gas pressures employed in MEMS vapor cells~\cite{rushton2023alignment}.

Hyperfine structure (HFS) techniques, such as coherent population trapping (CPT)~\cite{pollinger2018coupled,liang2014simultaneously} and direct microwave interrogation~\cite{aleksandrov2006magnetometer}, emerge as promising avenues for high scalar accuracy even in the challenging environment of MEMS cells. These techniques eliminate heading errors caused by NLZ effects by resolving Zeeman shifts between multiple hyperfine transitions. Compared to microwave interrogation, the all-optical functionality of CPT offers potential for miniaturization and reduced power consumption. However, CPT measurements tend to have worse sensitivity~\cite{batori2022mupop} and are also notably affected by light shifts. Thus far, despite the ability of both CPT and microwave interrogation techniques to function under high buffer gas pressures, there have been no studies examining the accuracy of HFS magnetometry within a MEMS cell.

A notable example of a microwave HFS OPM was built by E. B. Aleksandrov \textit{et. al.}~\cite{aleksandrov2006magnetometer} using continuous microwave interrogation of the $\sigma^{\pm}$ end-hyperfine resonances of $^{87}$Rb with unpolarized light. This sensor, operating with a cell volume of 63 cm$^3$, achieved 6-pT resolution over 0.1~s of measurement time, free from deadzones, and with heading errors within 0.5~nT, which were primarily attributed to residual magnetization in the sensor body. An alternative method to detect hyperfine resonances is to utilize the detuning-dependence of Rabi oscillation frequencies.  When combined with non-destructive measurement, such as Faraday rotation~\cite{kiehl2023coherence}, Rabi measurements could enable sensitive HFS magnetometry while also mitigating light shifts in the optical detection. This approach is also alluring because Rabi oscillation frequencies have recently been shown to enable self-calibrated vector magnetometry~\cite{thiele2018self}. Regardless of the method employed, a microwave HFS magnetometer operating over all magnetic field directions is prone to off-resonant driving that causes systematic shifts in the hyperfine resonances.  To date, detailed modeling of off-resonant driving in multilevel atomic systems, a critical factor for characterizing the ultimate accuracy limits of microwave HFS magnetometers, is lacking.

\begin{figure*}[!tbh]\centering
\includegraphics[width=.99\textwidth]{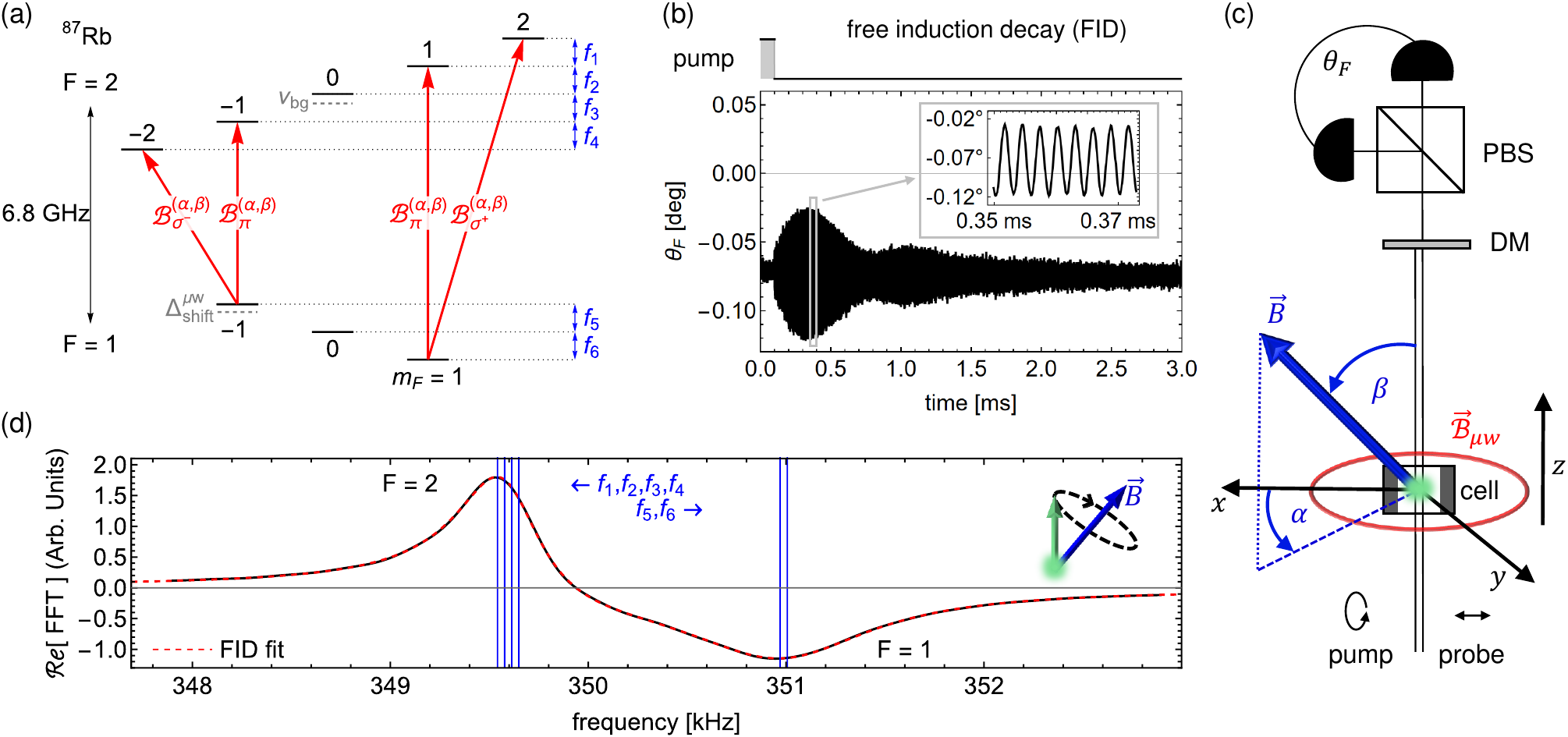}
\caption{(a) Energy-level diagram showing the four hyperfine transitions used in Rabi and Ramsey frequency spectroscopy (FS). Off-resonant $\mu\text{w}$ driving causes energy-level perturbations ($\Delta^{\mu\text{w}}_{\text{shift}}$) and buffer gas collisions cause a frequency shift in the hyperfine splitting ($\nu_{\text{bg}}$). (b) The FID pump timing diagram and the time-domain signal, measured at $\beta=106^{\circ}$ within the regime of low-spin polarization. In this regime, the FID signal contains both $F=1$ and $F=2$ precession frequencies. (c) Magnetometer apparatus to measure magnetic fields with the direction specified by azimuthal ($\alpha$) and polar ($\beta$) angles. An angled dichroic mirror (DM) reflects the pump. (d) Real part of the fast Fourier transform (FFT) of the FID signal shown in (b). Heading errors arise due to the uncertainty in the amplitudes and relative phases of the six Larmor precessional frequencies ($f_i$, blue lines).}
\label{fig:Figure1main}
\end{figure*}

In this work, we demonstrate microwave HFS magnetometry without systematic errors from off-resonant driving using Rabi and Ramsey frequency spectroscopy (FS) on the four hyperfine transitions of $^{87}$Rb shown in Fig.~\ref{fig:Figure1main}(a). With these two independent techniques we evaluate the heading error of a microfabricated
OPM based on free induction decay (FID) with subnanotesla accuracy. Importantly, both techniques perform \textit{in situ} measurements using the same vapor cell parameters, such as temperature and buffer gas pressure, as those used in sensitive FID measurements. This approach eliminates the characterization errors from magnetic field gradients, which often arise when employing spatially separate, high-accuracy magnetometers like Overhauser devices~\cite{duret1995overhauser} to benchmark heading error.

In Ramsey FS, systematic errors from off-resonant driving are avoided by using short periods of microwave interrogation within a $\pi/2-t_R-3\pi/2$ Ramsey sequence. Instead of detecting a central fringe, hyperfine resonances are measured by fitting to Ramsey fringes generated by varying the Ramsey time $t_R$ at several selected microwave detunings. This style of Ramsey interferometry is very robust against systematic errors, but has low sensitivity due to the many repeated Ramsey sequences that are required. Even so, we show that this approach  is useful as an accurate scalar reference to characterize heading errors.

In contrast, Rabi FS utilizes an atom-microwave Hamiltonian to accurately model the microwave detuning dependence of Rabi oscillation frequencies despite frequency shifts due to off-resonant driving. By using continuous nondestructive readout of spin-dynamics from the Faraday rotation of a far-detuned probe beam, this approach achieves sensitive measurements down to 80 pT$/\sqrt{\text{Hz}}$ limited by technical microwave noise. We further show with a Cram\'{e}r-Rao lower-bound (CRLB) analysis that the ultimate sensitivity of Rabi FS with our vapor cell parameters is 30 pT$/\sqrt{\text{Hz}}$. To ensure proper modeling of the atom-microwave coupling, we check the consistency between Rabi measurements driven by three distinct microwave polarization ellipses~\cite{thiele2018self}, which each induce unique frequency shifts due to off-resonant driving. 

We compare these two methods to FID measurements over a range of dc magnetic field directions at 50 $\mu$T. To prevent signal degradation in arbitrary magnetic field directions, both techniques employ adiabatic power ramps during optical pumping to suppress Larmor precession. We find that the Rabi and Ramsey techniques, despite their distinct concepts, both measure the FID heading error with agreement to within 0.6 nT. From theoretical simulations, we determine that the fundamental accuracy of both approaches is within 0.4 nT due to spin-exchange frequency shifts~\cite{micalizio2006spin,appelt1998theory}.

\section{Experimental Setup}

The magnetometer apparatus [Fig.~\ref{fig:Figure1main}(c)] consists of a $3\times3\times2$ mm$^3$ MEMS vapor cell with a single optical axis and filled with 180 Torr of N$_2$ buffer gas. The cell is contained inside a copper rectangular $(4.8\times4.8\times2\text{ cm}^3)$ microwave cavity that, through excitation of two linearly polarized modes, creates an arbitrarily shaped microwave polarization ellipse (MPE) in a plane orthogonal to the optical axis at the position of the atoms. This cavity is designed with a low quality factor of $Q=62$ (110-MHz linewidth) to minimize frequency dependence of the cavity modes. The cavity temperature is kept close to $100^\circ$C using joule heating, which in turn raises the temperature of the MEMS cell to a similar level. 

The MEMS cell and microwave cavity are housed within a three-dimensional (3D) coil system that generates a programmable 50-$\mu$T magnetic field $\vec{B}$, defined by azimuthal and polar angles $\alpha$ and $\beta$. This coil system defines an orthogonal reference frame $\mathcal{L}=(x,y,z)$, where a calibration corrects for misalignments between the coil pairs (see Appendix~\ref{sec:coilSystem}). Because heading error is predominantly influenced by the polar angle ($\beta$) between the pumping direction and the magnetic field $(\vec{B})$~\cite{lee2021heading}, we simplify this study, without loss of generality, by limiting the magnetic field orientations to the x-z plane ($\alpha=0$), except during the calibration of the coil system.

Along the cell optical axis, parallel to $\hat{z}$, propagates a 795-nm elliptically polarized pump beam, tuned within a few gigahertz of the $D_1$ line, and a 1-mW probe beam blue-detuned by 170 GHz from the 780-nm $D_2$ line. The pump frequency and polarization are tuned to depopulate the $F = 2$ manifold and enable strong Rabi signals across all hyperfine transitions, while still causing spin polarization for the FID measurement. Complete depopulation of the $F=2$ manifold is limited by the 5.6-GHz optical broadening from Rb-N$_2$ collisions. A polarimeter detects the Faraday rotation $\theta_F=g\langle S_z \rangle+\theta_0$ of the probe beam expressed in terms of the macroscopic z component of the electron spin, a coupling coefficient $g$, and an offset $\theta_0$~\cite{seltzer2008developments}.

\section{FID magnetometry}
For comparative demonstration, we study FID spin-precession signals with low atomic spin polarization where no accurate physical models for heading errors exist. To initiate FID measurements, as depicted in Fig.~\ref{fig:Figure1main}(b), a 100-$\mu$s pulse of pump light at 400 mW polarizes the atomic spins along the pump beam. In this low spin polarization regime, the FID spectrum [Fig.~\ref{fig:Figure1main}(d)] consists of both $F = 1$ and $F = 2$ Zeeman resonances that are separated at 50-$\mu$T by 1.4 kHz. The NLZ effect splits these resonances into frequency components $\{f_1,...,f_6\}$ separated by 36 Hz. We model the FID spectrum as two resonances $f_{L,\pm}(B)\approx \mu_B(g_s-g_i\pm 4g_i)B/4h$ that are the mean Zeeman splitting across the magnetic sublevels for the $F=I\pm1/2$ manifolds, where $I=3/2$ is the nuclear spin, $g_s$ and $g_i$ are the electronic and nuclear Land\'e $g$-factors, $h$ is Planck's constant, $\mu_B$ is the Bohr magneton, and $B$ is the magnetic field strength. The real component of the FID signal's Fourier transform in this model is given by
\begin{equation}
\label{eq:FSP}
Re[\text{FFT}] = \sum_{j=\pm}a_j\frac{\text{cos}(\phi_j)-\text{sin}(\phi_j)(f-f_{L,j})}{(f-f_{L,j})^2+w_j^2/4}
\end{equation}
where $\phi_{\pm}=2\pi f_{L,\pm} t_0\pm\phi/2$ are phase shifts due to a starting time offset $t_0$ with $\phi$ being a relative phase between the $I\pm1/2$ resonances respectively. Here the strength and broadening of this signal is given by amplitudes $a_{\pm}$ and linewidths $w_{\pm} \approx 1$ kHz. Based on the initial atomic state and the direction of $\vec{B}$, heading error arises in this model from the unresolved NLZ frequency components that bias the observed resonances from $f_{L,\pm}$.

\section{Microwave HFS magnetometry}
\label{sec:microwaveHFS}
We conduct microwave HFS magnetometry to circumvent these heading errors using either Rabi or Ramsey FS on the four hyperfine transitions diagrammed in Fig.~\ref{fig:Figure1main}(a). Because our pump beam generates spin polarization, as required for FID measurements, subsequent Larmor precession causes significant degradation of the Rabi and Ramsey signals. To circumvent this, we initialize the atomic ensemble in both methods with adiabatic optical pumping (AOP) to align the macroscopic atomic spin along the magnetic field $\vec{B}$. For each measurement, we execute AOP by first optically pumping the atomic ensemble for 50 $\mu$s at 100 mW and then linearly ramping off the pump power over the next 50 $\mu$s [see Fig.~\ref{fig:ramseyFig1}(a) and Fig.~\ref{fig:rabiMeasEigenvalue}(a) below]. A simple theoretical derivation for the spin alignment effect of AOP is discussed in Appendix~\ref{sec:aop}.

\subsection{Ramsey FS}

\begin{figure}[tbh]\centering
\includegraphics[width=.43\textwidth]{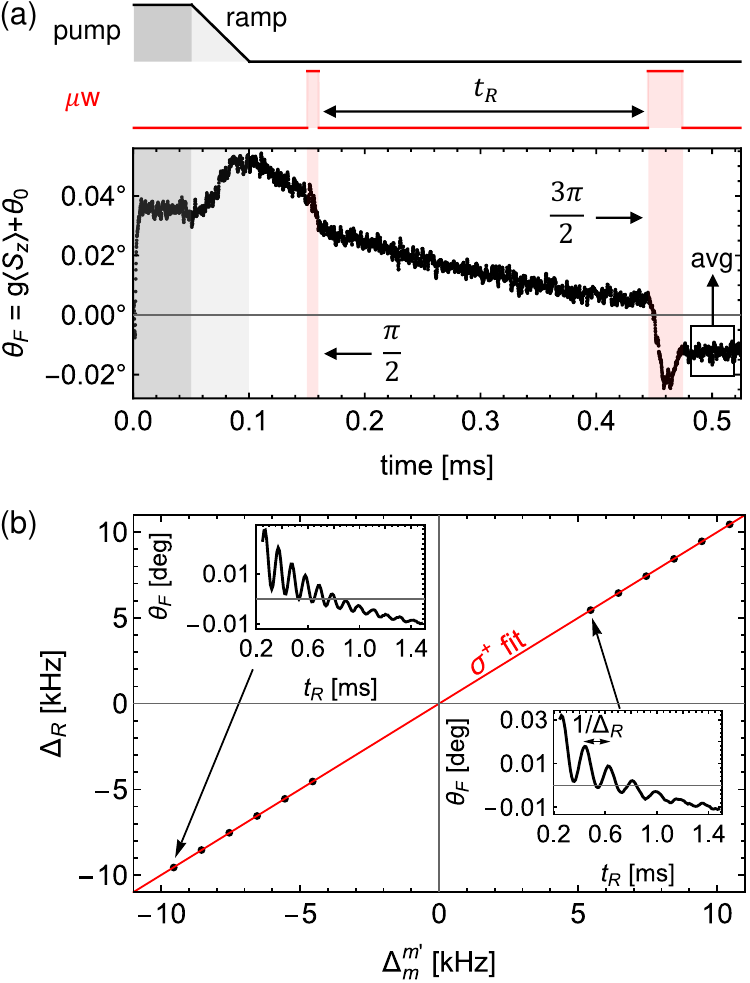}
\caption{(a) Faraday rotation signal during Ramsey interferometry and the associated timing diagram. (b) Ramsey frequency ($\Delta_R$) vs microwave detuning $(\Delta_m^{m^{\prime}})$ for the $\sigma^+$ transition. Insets show measured Ramsey fringes as a function of the Ramsey time $t_R$. A linear fit (red) extracts the transition resonance. Both (a) and (b) were measured at $\beta=34^{\circ}$.}
\label{fig:ramseyFig1}
\end{figure}

Ramsey FS utilizes pulsed microwave interrogation to accurately measure any hyperfine resonance $\nu_m^{m^{\prime}}\approx 6.8$ GHz between sublevels $\ket{1,m}$ and $\ket{2,m^{\prime}}$ without exact knowledge of off-resonant driving. This is achieved through a $\pi/2-t_R-3\pi/2$ Ramsey pulse sequence with $t_{\pi/2}=1/4\Omega_m^{m^{\prime}}\approx 10$ $\mu$s [Fig.~\ref{fig:ramseyFig1}(a)]. Satisfying this particular $t_{\pi/2}$ required manual adjustments to the microwave power at each magnetic field direction such that the Rabi frequency for each hyperfine transition satisfied $\Omega_m^{m^{\prime}}\approx 25$ kHz. Importantly, after each pulse sequence, we average the resulting Faraday signal for $50$ $\mu$s to filter out residual 350-kHz Larmor precession [Fig.~\ref{fig:ramseyFig1}(a)]. It should be emphasized that opting for a final $\pi/2$ pulse over a $3\pi/2$ pulse could marginally enhance the signal without affecting the systematic errors within the magnetometry protocol. The choice to use a $3\pi/2$ pulse in this study originates from its role in hyper-Ramsey interferometry, where it helps diminish systematic errors in detecting the central fringe~\cite{yudin2010hyper}, even though this specific scenario does not apply here.

\begin{figure}[tbh]\centering
\includegraphics[width=.5\textwidth]{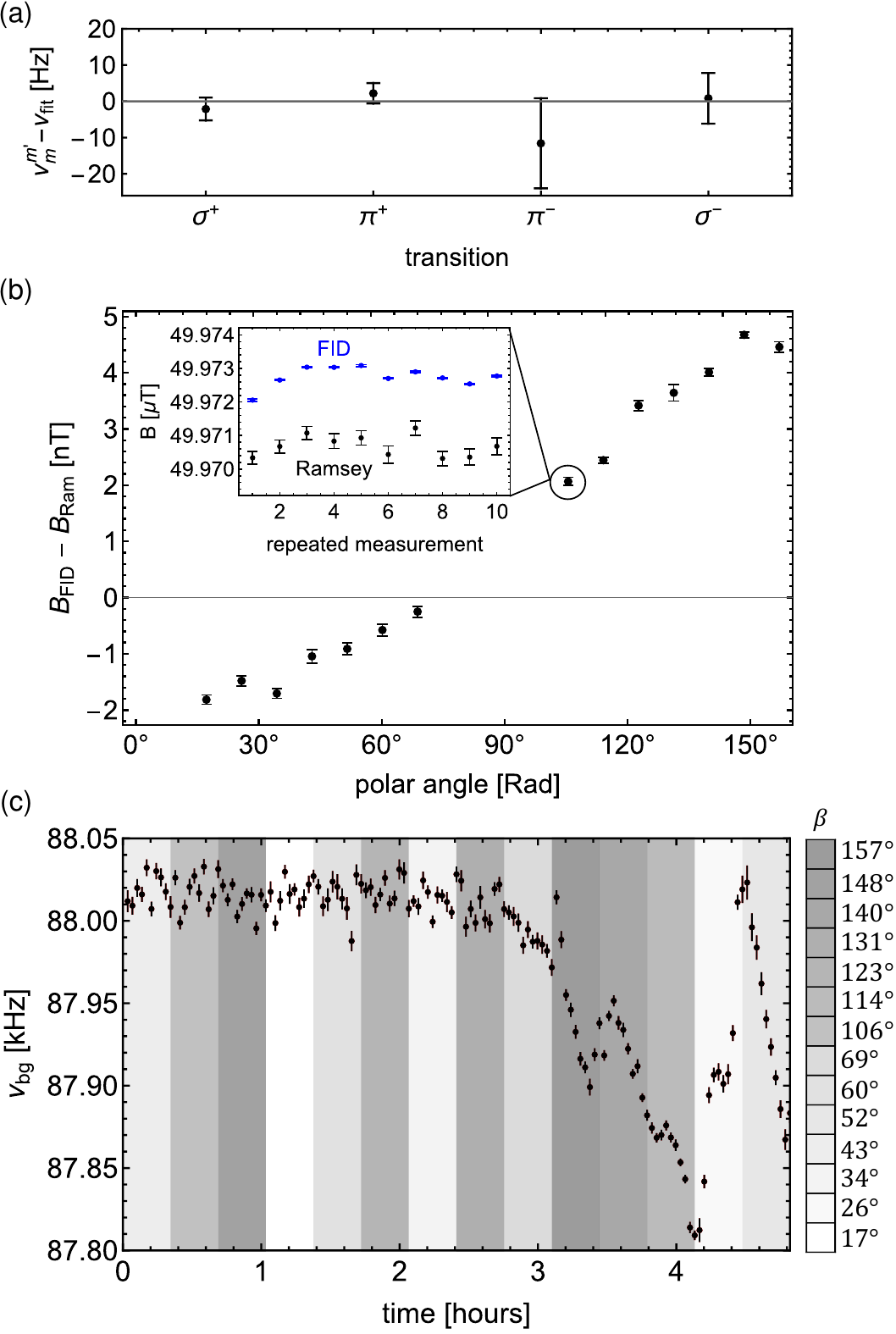}
\caption{Magnetic field strengths and pressure shifts evaluated over 14 different magnetic field orientations with Ramsey FS. Error bars show 68\% confidence interval. (a) Residuals of fitting Eq.~\eqref{eq:breitRabi} to the hyperfine resonances. The $\pi$ transitions between the $m_F=\pm 1$ sublevels are denoted as $\pi^{\pm}$. (b) Magnetic field strength residuals between FID Ramsey FS measurements over 14 different magnetic field orientations in the xz-plane. Inset shows repeated FID and Ramsey measurements.(c) The corresponding buffer gas pressure shifts ($\nu_{\text{bg}}$) obtained during Ramsey FS. The orientation of the magnetic field in these measurements is represented by the gray scale. Drift in $\nu_{\text{bg}}$, seen here, is due to a few $^{\circ}$C temperature drift of the microwave cavity.}
\label{fig:ramseyHeadErrOnly}
\end{figure}

We measure Ramsey fringes at each microwave detuning, as shown in the insets of Fig.~\ref{fig:ramseyFig1}(b), by varying the Ramsey free evolution time $t_R$ between 0.2 and 1.43 ms at 10-$\mu$s spacing. We fit these Ramsey fringes in the time domain with an exponentially decaying sinusoid
\begin{align}
\begin{split}
\label{eq:timeDomainFit}
\theta_F(t)=&a_0+a_{1}e^{-t/t_{1}}+a_{2}e^{-t/t_{2}}+\\&e^{-t/t_3} [a_3 \text{sin}(2\pi t f)+a_4 \text{cos}(2\pi t f)]
\end{split}
\end{align}
and force the fringe frequencies $f\rightarrow\Delta_R$ to be either positive or negative according to the sign of the microwave detuning $\Delta_m^{m^{\prime}}=\nu_{\mu \text{w}}-\nu_{m}^{m^{\prime}}$. In Eq.~\eqref{eq:timeDomainFit} the two dc-offset decay constants $t_1$ and $t_2$ are required to account for the atomic population redistribution arising from spin-exchange collisions~\cite{kiehl2023coherence}. Without influence from systematic shifts, $\Delta_R=\Delta_m^{m^{\prime}}$. We choose six microwave detunings $\Delta_m^{m^{\prime}}\in[5,10]$ kHz below and above each transition resonance as shown for the $\sigma^+$ transition in Fig.~\ref{fig:ramseyFig1}(b). All of these measurements are taken in random order to mitigate systematics from time-dependent drifts in the microwave field. By linear fitting $\Delta_R$ as a function of the microwave frequency $\nu_{\mu \text{w}}$, the x intercept measures $\nu_{m}^{m^{\prime}}$. The pressure shift, arising from N$_2$ buffer gas collisions $\nu_{\text{bg}}\approx 88$ kHz, and the magnetic field strength $B\approx 50$ $\mu$T are obtained by fitting $\nu_{m}^{m^{\prime}}$ measurements [Fig.~\ref{fig:ramseyHeadErrOnly}(a)] to
\begin{equation}
\label{eq:breitRabi}
\frac{h\nu_{m}^{m^{\prime}}}{\Delta E}=\frac{m^{\prime}-m}{g_s/g_i-1}x+\frac{1}{2}\sum_{M=m,m^{\prime}}\sqrt{1+\frac{4Mx}{2I+1}+x^2}
\end{equation}
where $x=(g_s-g_i)\mu_B B/\Delta E$ and $\Delta E=(A+h\nu_{\text{bg}}/2)(I+1/2)$ is the hyperfine splitting expressed in terms of the magnetic dipole hyperfine constant $A$. Differences between scalar measurements using FID and Ramsey FS, each repeated 10 times at 14 magnetic field directions, are plotted in Fig.~\ref{fig:ramseyHeadErrOnly}(b). These differences, presumably from FID heading error, are contained within $5$ nT.

All Ramsey FS measurements shown in Fig.~\ref{fig:ramseyHeadErrOnly}(b) took a total of 4.8 hours, attributed to the extensive repetition of Ramsey sequences to vary $t_R$ and the required measurement downtime to keep the cavity temperature close to 100$^{\circ}$C. This downtime could be significantly shortened using magnetically quiet cell-heating methods such as laser heating~\cite{mhaskar2012low}. Along with degrading sensitivity, this several-hour measurement period makes Ramsey FS susceptible to errors arising from apparatus drifts. One form of drift is the temperature instability in our microwave cavity, varying by a few degrees Celsius. This drift is inferred from the pressure shift ($\nu_{\text{bg}}$) variations, as depicted in Fig.~\ref{fig:ramseyHeadErrOnly}(c). Additional information behind the Ramsey measurement sequence is provided in Appendix~\ref{sec:measSeq}.

\subsection{Rabi FS}
\begin{figure}[tbh]\centering
\includegraphics[width=.43\textwidth]{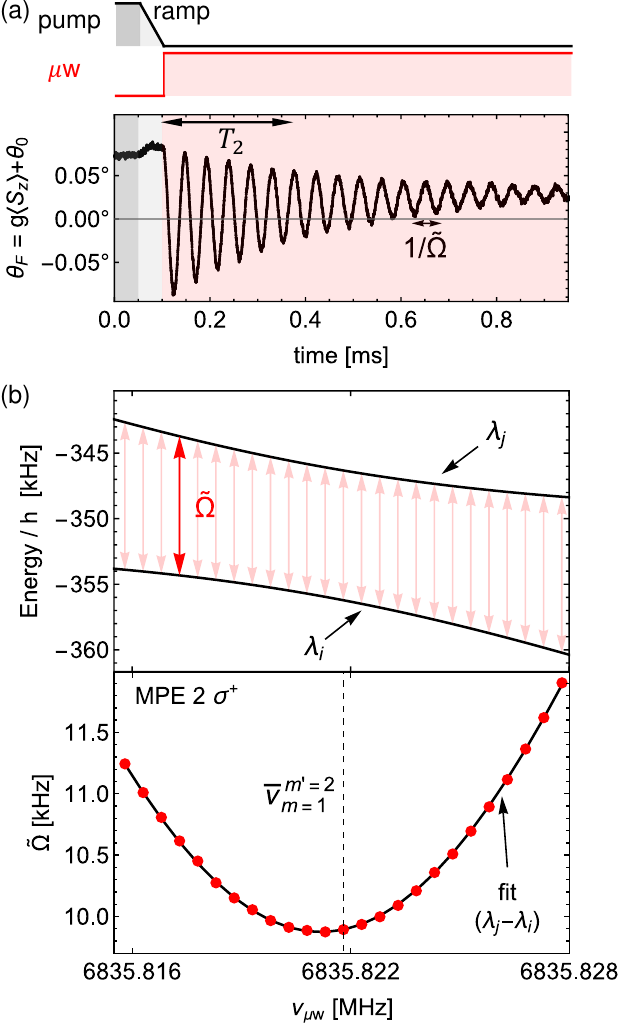}
\caption{(a) Rabi oscillation timing diagram. (b) $\sigma^+$ Rabi measurements and corresponding eigenvalue differences $\lambda_j-\lambda_i$ of $H$ (Eq.~\eqref{eq:Hamiltonian}) versus the microwave frequency ($\nu_{\mu\text{w}}$).}
\label{fig:rabiMeasEigenvalue}
\end{figure}

Rabi FS [see Fig.~\ref{fig:rabiMeasEigenvalue}] measures hyperfine resonances from the detuning dependence of generalized Rabi frequencies ($\tilde{\Omega}_m^{m^{\prime}}$) that are fitted from Rabi oscillations using Eq.~\eqref{eq:timeDomainFit}. A low-pass filter, discussed in Appendix~\ref{sec:RabiFiltering} is applied to each Rabi oscillation to eliminate residual 350-kHz Larmor precession. The generalized Rabi frequency ($\tilde{\Omega}_m^{m^{\prime}}$) detuning dependence ($\Delta_m^{m^{\prime}}$) is given within a two-level formalism by
\begin{equation}
\label{eq:2lvl}
\tilde{\Omega}_m^{m^{\prime}}\approx \sqrt{(\Delta_m^{m^{\prime}})^2+|\Omega_m^{m^{\prime}}|^2}.
\end{equation}
In contrast to Ramsey FS, this approach demands precise modeling of atom-microwave coupling to accommodate for frequency shifts from off-resonant driving in the multilevel atomic system. To incorporate the effects of off-resonant driving, the generalized Rabi frequency ($\tilde{\Omega}_m^{m^{\prime}}$) is modeled in terms of dressed state energies $\lambda_j$ and $\lambda_i$ that correspond to the pair of states coupled by the microwave field. This relationship is expressed as
\begin{equation}
\label{eq:rabiLambda}
\tilde{\Omega}_m^{m^{\prime}}= \delta \lambda_m^{m^{\prime}} \equiv (\lambda_j-\lambda_i)/h
\end{equation}
where $\lambda_j$ and $\lambda_i$ are eigenvalues of the atom-microwave Hamiltonian 
\begin{align}
\begin{split}
\label{eq:Hamiltonian}
H=&\mathcal{M}\big[(A+h\frac{\nu_{\text{bg}}}{2})\mathbf{S}\cdot\mathbf{I}+\mu_B(g_sS_z+g_iI_z)B\big]\mathcal{M}^{\dagger}\\ -I_2& h\nu_{\mu\text{w}}+\sum_{|m-m^{\prime}|\leq 1}\frac{h}{2}\Big[\ket{\overline{2,m^{\prime}}}\Omega_{m}^{m^{\prime}}\bra{\overline{1,m}}+\text{H.c.}\Big].
\end{split}
\end{align}

\begin{figure*}[!tbh]\centering
\includegraphics[width=.99\textwidth]{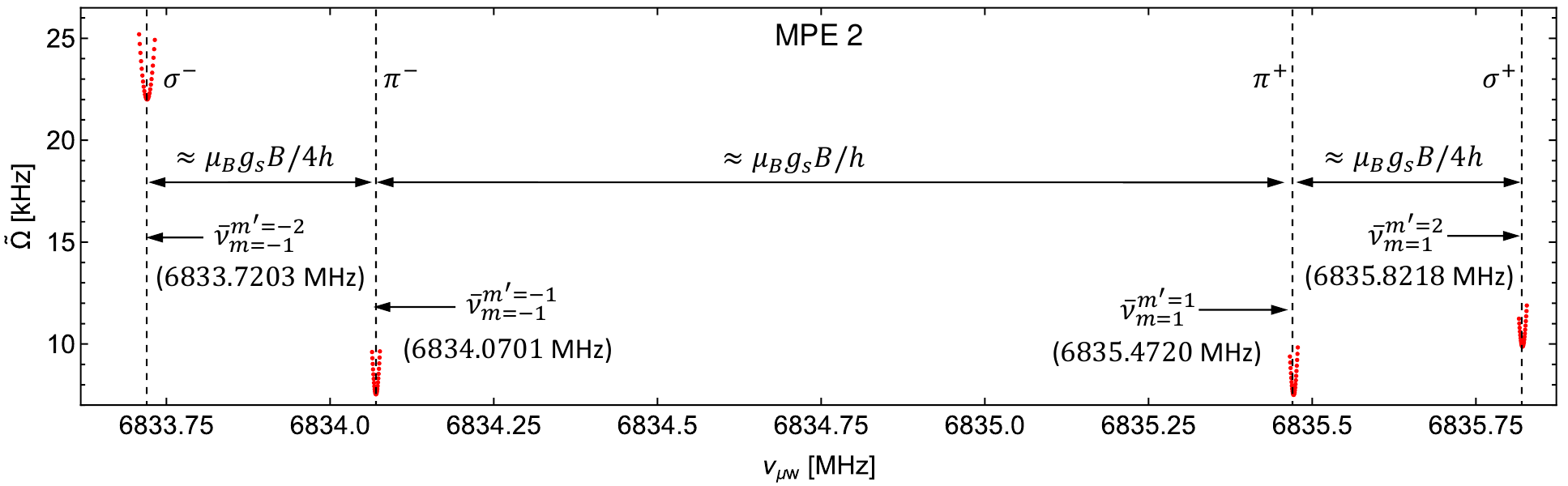}
\caption{Generalized Rabi frequency measurements at $\beta=34^{\circ}$, driven by MPE 2, with center microwave frequencies $\overline{\nu}_m^{m^{\prime}}$ (dashed lines). The $\pi$ transitions between the $m_F=\pm 1$ sublevels are denoted as $\pi^{\pm}$ [see Fig.~\ref{fig:rabMagDifferentTrans}].}
\label{fig:rabiFreqSpecn2}
\end{figure*}

Details on how to choose the eigenvalue pair ($\lambda_j,\lambda_i$) matching a specific $\tilde{\Omega}_m^{m^{\prime}}$ can be found in Appendix~\ref{sec:EigenvalueSelect}. Hamiltonian~\eqref{eq:Hamiltonian} is defined within an atom frame $\mathcal{A}=(x_a,y_a,z_a)$ where the $z_a$ direction is aligned with the magnetic field direction $(\alpha,\beta)$ with respect to the lab frame $\mathcal{L}$ and characterized by electronic $\mathbf{S}=(S_x,S_y,S_z)$ and nuclear $\mathbf{I}=(I_x,I_y,I_z)$ spin operators defined in the $\ket{F,m_F}$ basis. In order to preserve NLZ effects during the rotating-wave approximation (RWA), we work in a modified hyperfine basis $\ket{\overline{F,m}}=\mathcal{M}\ket{F,m}$ with $\mathcal{M}$ being defined as the operator that diagonalizes the hyperfine and Zeeman terms in the first line of Eq.~\eqref{eq:Hamiltonian}. 

The second line of Eq.~\eqref{eq:Hamiltonian} describes the atom-microwave coupling. Here, the energy difference between the $F=1,2$ manifolds is reduced by the microwave frequency $(\nu_{\mu\text{w}})$ during the RWA, where $I_2$ is the identity operator for the $F=2$ manifold. The Rabi frequency ($\Omega_m^{m^{\prime}}$), characterizing the atom-microwave coupling strength between sublevels $\ket{1,m}$ and $\ket{2,m^{\prime}}$, is given in terms of the spherical microwave component ($\mathcal{B}_k^{(\alpha,\beta)}$) and the magnetic transition dipole moment ($\mu_m^{m^{\prime}}$) through
\begin{equation}
\label{eq:rabiFreq}
\Omega_m^{m^{\prime}}=\mu_m^{m^{\prime}}\mathcal{B}^{(\alpha,\beta)}_k/h.
\end{equation}
Here, $k=\pm,\pi$ denotes the polarization of the hyperfine transition. Appendix~\ref{sec:transDipoles} contains additional details on $\mu_m^{m^{\prime}}$ and the transformation matrix $\mathcal{M}$. The spherical microwave components $\mathcal{B}_k^{(\alpha,\beta)}$ are defined within the atom frame ($\mathcal{A}$) as
\begin{equation}
\label{eq:sphericalMic}
\mathcal{B}_k^{(\alpha,\beta)}=R_y(-\beta)R_z(-\alpha)\vec{\mathcal{B}}\cdot\epsilon_k
\end{equation}
where $\vec{\mathcal{B}}=(\mathcal{B}_xe^{-i\phi_x},\mathcal{B}_ye^{-i\phi_y},\mathcal{B}_z)$ is a complex phasor written in terms of microwave amplitudes $(\mathcal{B}_x,\mathcal{B}_y,\mathcal{B}_z)$ and relative phases $(\phi_x,\phi_y)$ defined in the lab frame $\mathcal{L}$. The spherical projection operators $\epsilon_\pm= \{\frac{1}{\sqrt{2}},\mp \frac{i}{\sqrt{2}},0\}$ and $\epsilon_{\pi}=\{0,0,1\}$ are also defined within $\mathcal{A}$. In Eq.~\eqref{eq:sphericalMic} the phasor $\vec{\mathcal{B}}$ is rotated into $\mathcal{A}$ through 3D rotation operators $R_{y,z}$ that are defined with respect to the lab frame $\mathcal{L}$. This phasor describes any microwave field through \begin{equation}
\label{eq:MPE}
\vec{\mathcal{B}}(t)=\frac{1}{2}[\vec{\mathcal{B}}e^{-i\omega_{\mu\text{w}}t}+\vec{\mathcal{B}}^{*}e^{i\omega_{\mu\text{w}}t}].
\end{equation}

\begin{figure}[tbh]\centering
\includegraphics[width=.45\textwidth]{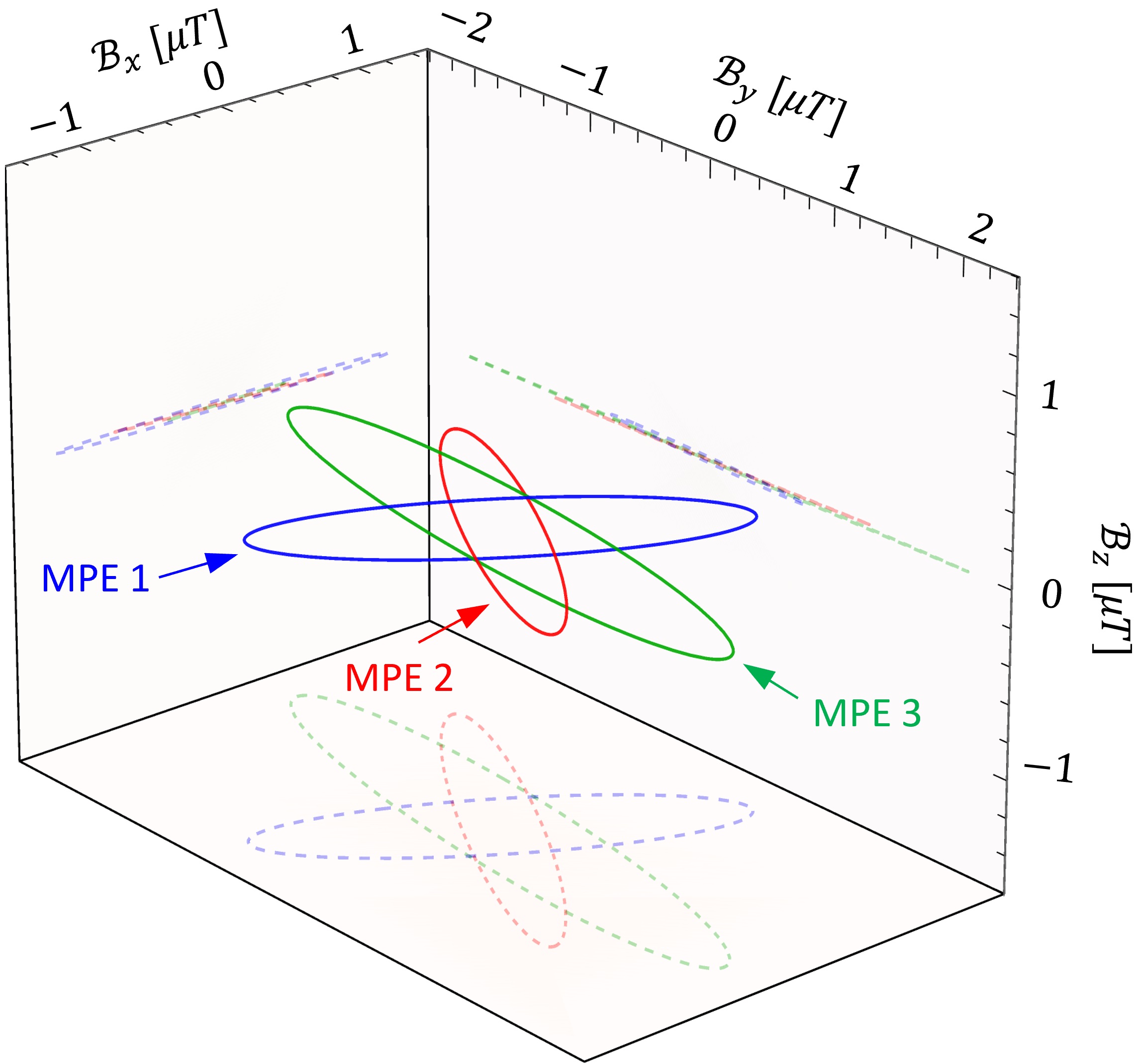}
\caption{Calibrated microwave polarization ellipses (MPEs) used for Rabi FS. For further details on these calibrations, see Appendix~\ref{sec:MPECharacterization}.}
\label{fig:calibratedMPEs}
\end{figure}

\begin{figure*}[!tbh]\centering
\includegraphics[width=.90\textwidth]{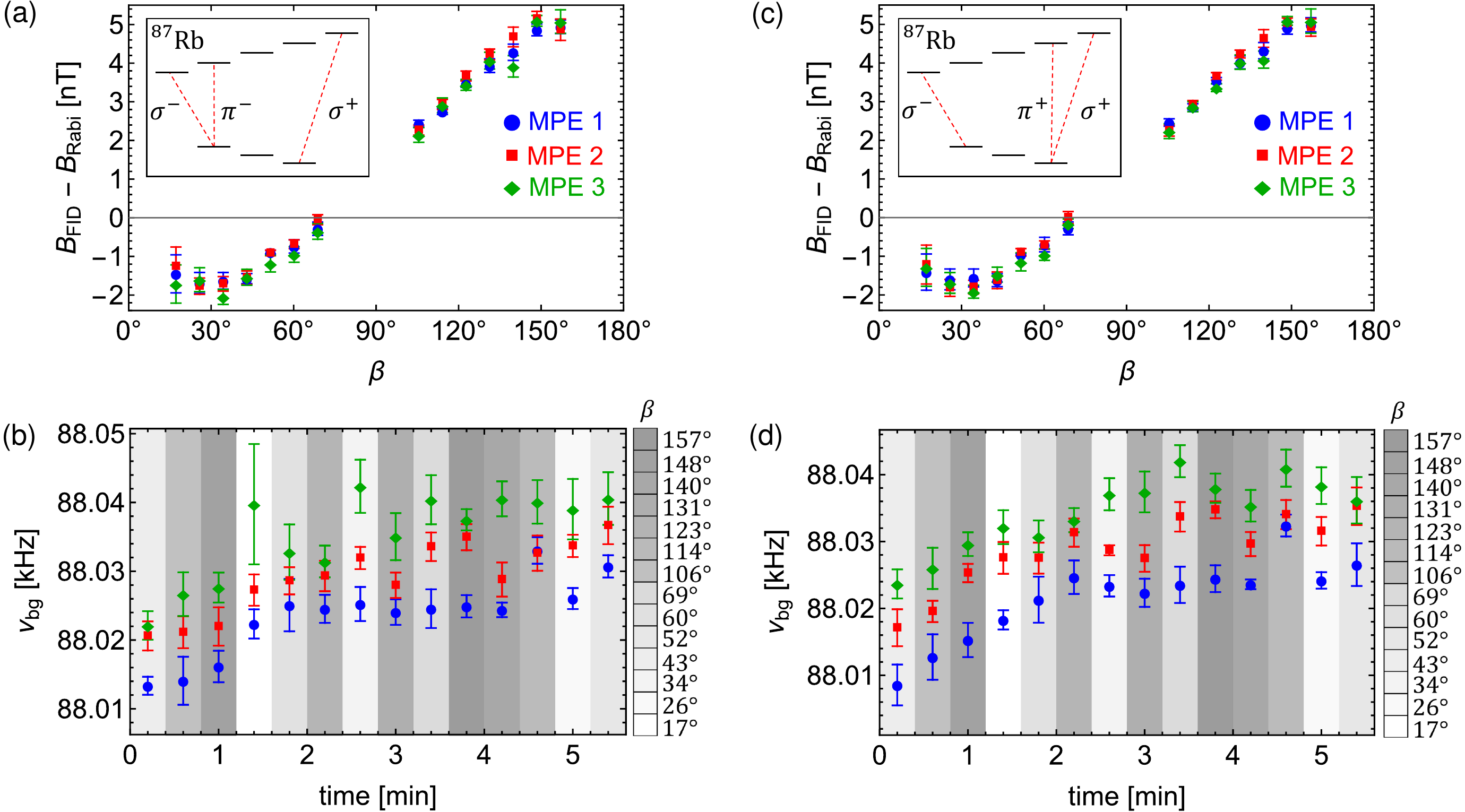}
\caption{Magnetic field strength ($B$) and pressure shift ($\nu_{\text{bg}}$) measurements using Rabi FS with (a,b) $(\sigma^-,\pi^-,\sigma^+)$ transitions and (c,d) $(\sigma^-,\pi^+,\sigma^+)$ transitions. Drift in the $\nu_{\text{bg}}$ is attributed to small temperature drifts of the vapor cell.}
\label{fig:rabMagDifferentTrans}
\end{figure*}

With the $\delta \lambda_m^{m^{\prime}}$ model given in Eq.~\eqref{eq:rabiLambda}, the magnetic field strength ($B$), the pressure shift ($\nu_{\text{bg}}$), and the three spherical microwave components ($\mathcal{B}_{\sigma^{\pm}}^{(\alpha,\beta)}$ and $\mathcal{B}_{\pi}^{(\alpha,\beta)}$), assumed to be positive, are fitted from generalized Rabi frequency measurements about either the $(\sigma^{-},\pi^{-},\sigma^+)$ or the $(\sigma^{-},\pi^{+},\sigma^+)$
transitions. During these fits the generalized Rabi frequencies are weighted in terms of the fitting error deduced from the time-domain fits with Eq.~\eqref{eq:timeDomainFit}. Attempting to fit the $\pi^{\pm}$ hyperfine transitions with a single $\mathcal{B}_{\pi}^{(\alpha,\beta)}$ parameter simultaneously will lead to errors, as the spherical microwave components depend on the microwave frequency. This issue stems from the 110-MHz linewidth of the microwave cavity modes, along with residual frequency dependence within the filters, amplifiers, and coaxial cables that comprise the microwave electronics.

Importantly, the assumption that the spherical microwave components are positive values, rather than complex, eliminates the need for an independent calibration of the microwave field amplitudes and phases, as well as for determining the direction of the magnetic field, $(\alpha,\beta)$. This assumption is valid for the weak coupling ($|\Omega_m^{m^{\prime}}|\ll f_{L,\pm}$) employed in this experiment.

As shown in Fig.~\ref{fig:rabiFreqSpecn2}, Rabi measurements about each transition are driven at 25 microwave detunings ($\Delta_m^{m^{\prime}}$) spaced by 800 Hz, with center frequency $\overline{\nu}_{m}^{m^{\prime}}$. We conducted three sets of Rabi measurements, each with distinct microwave cavity mode excitations. The resulting microwave fields are defined by the MPEs depicted in Fig.~\ref{fig:calibratedMPEs}. Because of the unique amount of off-resonant driving caused by each MPE, the use of multiple MPEs verifies the accuracy of Eq.~\eqref{eq:rabiLambda} from the consistency in the fitted values of $B$ and $\nu_{\text{bg}}$ across these independent data sets. Additional information behind the Rabi measurement sequence and characterization of the MPEs for the purposes of modeling potential systematic errors are discussed in Appendices~\ref{sec:measSeq} and \ref{sec:MPECharacterization}, respectively.   

The discrepancy between the magnetic field strengths of FID and Rabi measurements are shown in Fig.~\ref{fig:rabMagDifferentTrans} for fits using generalized Rabi frequencies about either the $(\sigma^-,\pi^+,\sigma^+)$ or the $(\sigma^-,\pi^-,\sigma^+)$ transitions. The FID discrepancy shows the same qualitative behavior as found using Ramsey FS in Fig.~\ref{fig:ramseyHeadErrOnly}(b). Furthermore, the pressure shift measurements, up to effects from vapor temperature drifts, are consistent between Rabi and Ramsey FS. Vapor temperature drifts during the 5.5-min measurement duration are attributed to the cause of the approximate $20$-Hz drift in the pressure shift in Fig.~\ref{fig:rabMagDifferentTrans}(b,d). Despite general agreement between magnetic field strength measurements of different MPEs, there are approximately $10$-Hz descrepancies between the pressure shift measurements of different MPEs. This discrepancy is consistent with predicted systematic errors arising from the frequency dependence of the MPE parameters discussed in Appendix~\ref{sec:ScalarTheoritcalSim}.

\section{Sensitivity Analysis of Rabi FS and FID}
\begin{figure}[!tbh]
\begin{center}
\includegraphics[width=.45\textwidth]
{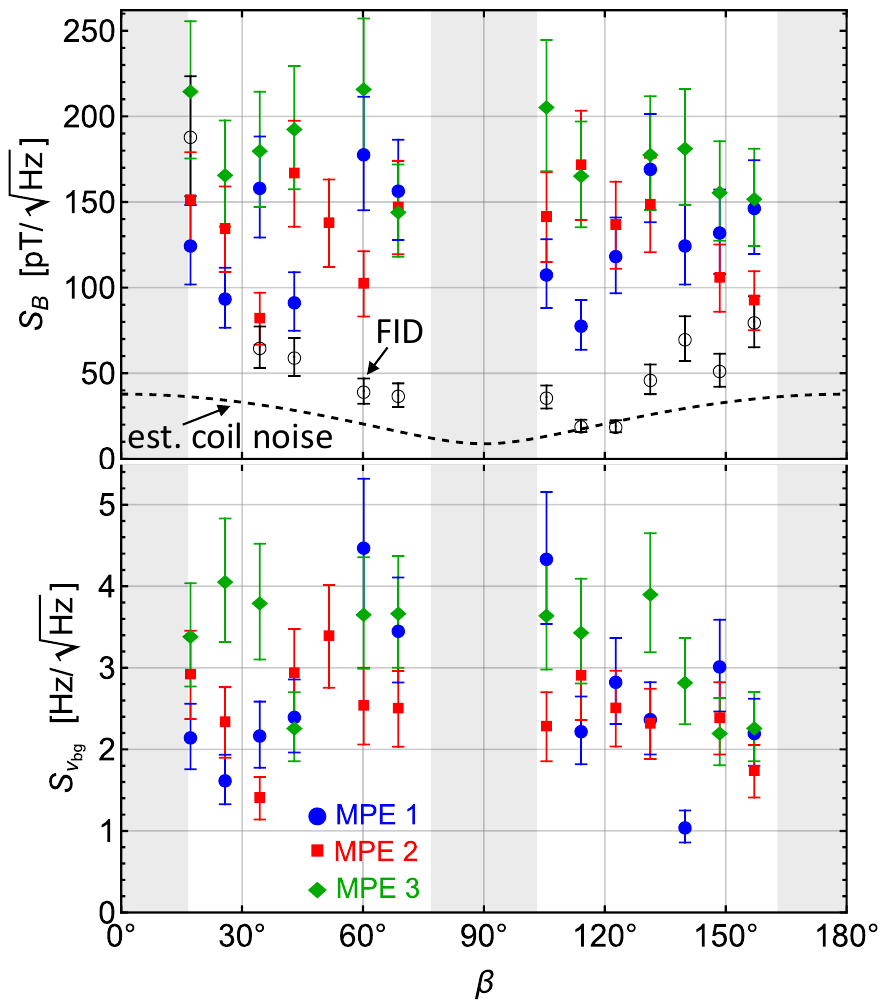}
\end{center}
\caption{Magnetic field strength ($S_B$) and pressure shift ($S_{\nu_{\text{bg}}}$) sensitivities using Rabi FS. The FID sensitivities (black circles) and the expected coil noise (black dashed), assessed at our 0.3 Hz measurement repetition rate, are overlaid for comparison. Error bars denote a 95\% confidence interval due to the fact that the sensitivities are calculated from the standard error ($\sigma_B$) of only eight repeated measurements.}
\label{fig:rabiSensitivity}
\end{figure}

The magnetic field strength and pressure shift sensitivities for different $\beta$, as measured with Rabi FS, are reported in Fig.~\ref{fig:rabiSensitivity}, with the best sensitivities observed to be near $80$ pT/$\sqrt{\text{Hz}}$ and $1.0$ Hz/$\sqrt{\text{Hz}}$. The magnetic field strength sensitivity~\cite{hunter2018free} 
\begin{equation}
\label{eq:magneticSensitivity}
S_B=\sigma_B\sqrt{2t_m}
\end{equation}
is calculated from the standard error ($\sigma_B$) of eight repeated measurements of the magnetic field strength ($B$) over a total measurement time $t_m=8\times100$ ms. The sensitivity $S_{\nu_{\text{bg}}}$ of the buffer gas pressure shift is calculated similarly. Downtime between measurements, reserved to maintain the cavity and cell temperature through joule heating [see Appendix~\ref{sec:measSeq}], is not included in the active measurement time $t_m$ for calculating Eq.~\eqref{eq:magneticSensitivity}.

Also shown in Fig.~\ref{fig:rabiSensitivity} is the estimated magnetic noise in our coil system [see Appendix~\ref{sec:coilSystem}] and the corresponding scalar sensitivities from eight repeated FID measurements that are evaluated from an active measurement time $t_m=8\times9$ ms. Each FID scalar value is derived from a 9-ms FID train consisting of three FIDs each lasting 3 ms. Error bars in Fig.~\ref{fig:rabiSensitivity} denote the 95\% confidence interval from the uncertainty in the standard error ($\sigma_B$) calculated from only eight measurements.

To assess the inherent precision of Rabi FS, we first determine the minimum variance $\sigma_{\Omega}^2$ (CRLB) for the frequency uncertainty of a Rabi oscillation. The CRLB for a damped sinusoid is given by~\cite{gemmel2010ultra,grujic2015sensitive} 
\begin{equation}
\sigma_{\Omega}^2\geq \frac{12}{(2\pi)^2 (A_{\theta}/\sigma_{\theta})^2 f_s T_r^3}C
\end{equation}
where $A_{\theta}/\sigma_{\theta}$ is the signal-to-noise ratio (SNR), $T_r=0.85$ ms is the measurement time, and $f_s=10$ MHz is the sampling rate. Factor $C$ is an overall constant given in terms of the dephasing time $\gamma_2=1/T_2$ and the number of samples $N=f_s T_r=8500$:
\begin{equation}
C=\frac{N^3}{12}\frac{(1-z^3)^3(1-z^{2N})}{z^2(1-z^{2N})^2-N^2z^{2N}(1-z^2)^2}
\end{equation}
with $z=e^{-\gamma_2/f_s}$.

Taking the MPE-2 Rabi measurements at $\beta=42.4^{\circ}$ as an example, the mean near-resonant Rabi oscillation amplitude across all four hyperfine transitions is $A_{\theta}=0.042^{\circ}$, with a dephasing time of $T_2=0.3$ ms. From these parameters and the polarimeter noise $\sigma_{\theta}=0.0043^{\circ}$, discussed in Appendix~\ref{sec:polarimeterNoise}, the CRLB standard error for a near-resonant Rabi oscillation is $\sigma_{\Omega} = 1.1$ Hz. For Rabi oscillations detuned by ($\Delta_m^{m^{\prime}}$), we model the frequency uncertainty ($\sigma_{\tilde{\Omega}}$) with
\begin{equation}
\sigma_{\tilde{\Omega}}=\sigma_{\Omega}\frac{(\Omega_m^{m^{\prime}})^2+(\Delta_m^{m^{\prime}})^2}{(\Omega_m^{m^{\prime}})^2},
\end{equation}
as expected from the population dynamics of Rabi oscillations in a two-level system.

From the CRLB uncertainty ($\sigma_{\tilde{\Omega}}$) of a generalized Rabi frequency, we determine the uncertainty ($\sigma_{\nu}$) in the hyperfine transition resonance $\nu_m^{m^{\prime}}$, based on 25 Rabi oscillations with detunings $(\nu-\nu_m^{m^{\prime}})=j\Omega/12$, where $j$, an integer, ranges between $-12$ and $12$. We find that the frequency uncertainty in the transition resonance $\nu_m^{m^{\prime}}$ is $\sigma_{\nu}/\sigma_{\Omega}\approx 0.65$ from repeated fits of fake Rabi data, calculated from Eq.~\eqref{eq:2lvl}, with added Gaussian noise characterized by the variance $\sigma_{\tilde{\Omega}}^2$. This result is independent of the Rabi rate ($\Omega_m^{m^{\prime}}$) because of how the detunings ($\Delta_m^{m^{\prime}}$) are assumed to scale with $\Omega_m^{m^{\prime}}$.

To avoid unnecessary complexity in the CRLB analysis, we consider Rabi oscillations only about the $\sigma^{\pm}$ transitions. We obtain the CRLB uncertainty for the magnetic field strength ($\sigma_B$) and the pressure shift ($\sigma_{\nu_{\text{bg}}}$) by subtracting and adding the $\sigma^{\pm}$ transition resonances and propagating the hyperfine resonance uncertainty $\sigma_{\nu}$, namely,
\begin{align}
&(\nu_{m=1}^{m^{\prime}=2}-\nu_{m=-1}^{m^{\prime}-2}) \pm\sqrt{2}\sigma_{\nu}\approx 6\gamma (B\pm \sigma_B)\\
&(\nu_{m=1}^{m^{\prime}=2}+\nu_{m=-1}^{m^{\prime}-2}) \pm\sqrt{2}\sigma_{\nu} \approx2(\nu_{\text{hfs}}+\nu_{\text{bg}}\pm\sigma_{ \nu_{\text{bg}}}).
\end{align}
Here $\gamma\approx7$ Hz/nT is the gyromagnetic ratio and the unperturbed hyperfine frequency $\nu_{\text{hfs}}$ is taken to have no uncertainty.  Assuming a total measurement time of $t_m=50$ ms to make all 50 $\sigma^{\pm}$ Rabi oscillation measurements, these CRLB uncertainties are converted to sensitivities by
\begin{align}
\label{eq:magIdealSens2}
S_B=\sigma_B\sqrt{2t_m}=\sigma_{\Omega}\frac{0.65}{3 \gamma}\sqrt{t_m}=\text{30 pT}/\sqrt{\text{Hz}}\\
S_{\nu_{\text{bg}}}=\sigma_{\nu_{\text{bg}}}\sqrt{2t_m}=\sigma_{\Omega}0.65\sqrt{t_m}=\text{0.62 Hz}/\sqrt{\text{Hz}}.
\label{eq:pShiftIdealSens2}
\end{align}

In addition to current noise in the coil system, we attribute microwave field drift to why the measurements in Fig.~\ref{fig:rabiSensitivity} are a few factors higher than the CRLB sensitivity limits in Eqs.~\eqref{eq:magIdealSens2} and \eqref{eq:pShiftIdealSens2}. Another factor that limits the sensitivity in our Rabi measurements is imperfect state preparation into the $F=1$ manifold. This limitation arises from optical broadening due to buffer gas collisions, which hinders the complete depopulation of the $F=2$ manifold. If the atomic population was initialized equally among the $F=1$ manifold the CRLB sensitivity limits, that assume measured Rabi oscillation SNR, would improve by a few factors.

For completeness, we also apply the CRLB sensitivity analysis to our FID measurements, as done in Refs.~\cite{grujic2015sensitive,hunter2018free}. A notable feature of our FID signals, arising from significant $F=1$ atomic population, is that they contain two frequency components corresponding to the spin precession in both of the $F=1$ and $F=2$ hyperfine manifolds [see Fig.~\ref{fig:Figure1main}(d)]. We determine the overall magnetic sensitivity ($S_B$) using the sensitivities of the individual FID frequency components, $S_{B,1}$ and $S_{B,2}$, as
\begin{equation}
\label{eq:CRLBFIDSensitivity}
S_B=\frac{1}{1/S_{B,1}+1/S_{B,2}}.
\end{equation}
The sensitivities $S_{B,i}=\gamma\sigma_L$ are calculated from the CRLB variance ($\sigma_L^2$) in the Larmor precession frequency and the gyromagnetic ratio $\gamma$. For the FID measurements at $\beta=106^{\circ}$, the amplitude ($A_{\theta}$) and dephasing time ($T_2$) are $0.033^{\circ}$ and $0.51$ ms for the $F=2$ component, and $0.039^{\circ}$ and $0.26$ ms for the $F=1$ component, respectively. Dephasing times are estimated from the linewidths in Eq.~\eqref{eq:FSP} through $T_2=(\pi w_{j})^{-1}$. With the measurement time for a single FID being $T_r=3$ ms, these parameters correspond to a CRLB uncertainty $\sigma_{L}=2.3$ Hz ($S_{B,2}=25$ pT$/\sqrt{\text{Hz}}$) and $\sigma_{L}=5.4$ Hz ($S_{B,1}=61$ pT$/\sqrt{\text{Hz}}$) for the $F=2$ and $F=1$ frequency components respectively. From Eq.~\eqref{eq:CRLBFIDSensitivity}, we estimate the CRLB magnetic sensitivity for this FID measurement to be $S_B=18$ pT$/\sqrt{\text{Hz}}$.

A summary of the CRLB sensitivity analysis for both the Rabi and FID measurements is shown in Table~\ref{tab:CRLBRabiScalar}.

\begin{table}[!th]
\caption{\label{tab:CRLBRabiScalar}
CRLB sensitivitiy estimates for Rabi FS and FID measurements at polar angle $\beta$ using measured signal amplitudes ($A_{\theta}$) and dephasing rates ($T_2$). All estimates assume a Faraday rotation noise floor of $\sigma_{\theta}=0.0043^{\circ}$, as characterized in Appendix~\ref{sec:polarimeterNoise}.}
\begin{center}
\begin{tabular}{|c||c|c|}
\hline
\textrm{Measurement type} &
$S_B$ (pT$/\sqrt{\text{Hz}}$)&
$S_{\nu_{\text{bg}}}$ (Hz$/\sqrt{\text{Hz}}$)\\
\hline
Rabi ($\beta=42.4^{\circ}$)&30&0.62\\
FID ($\beta=106^{\circ}$)&18&-----\\
\hline
\end{tabular}
\end{center}
\end{table}
\section{Accuracy Analysis}

\begin{figure}[!tbh]
\begin{center}
\includegraphics[width=.47\textwidth]
{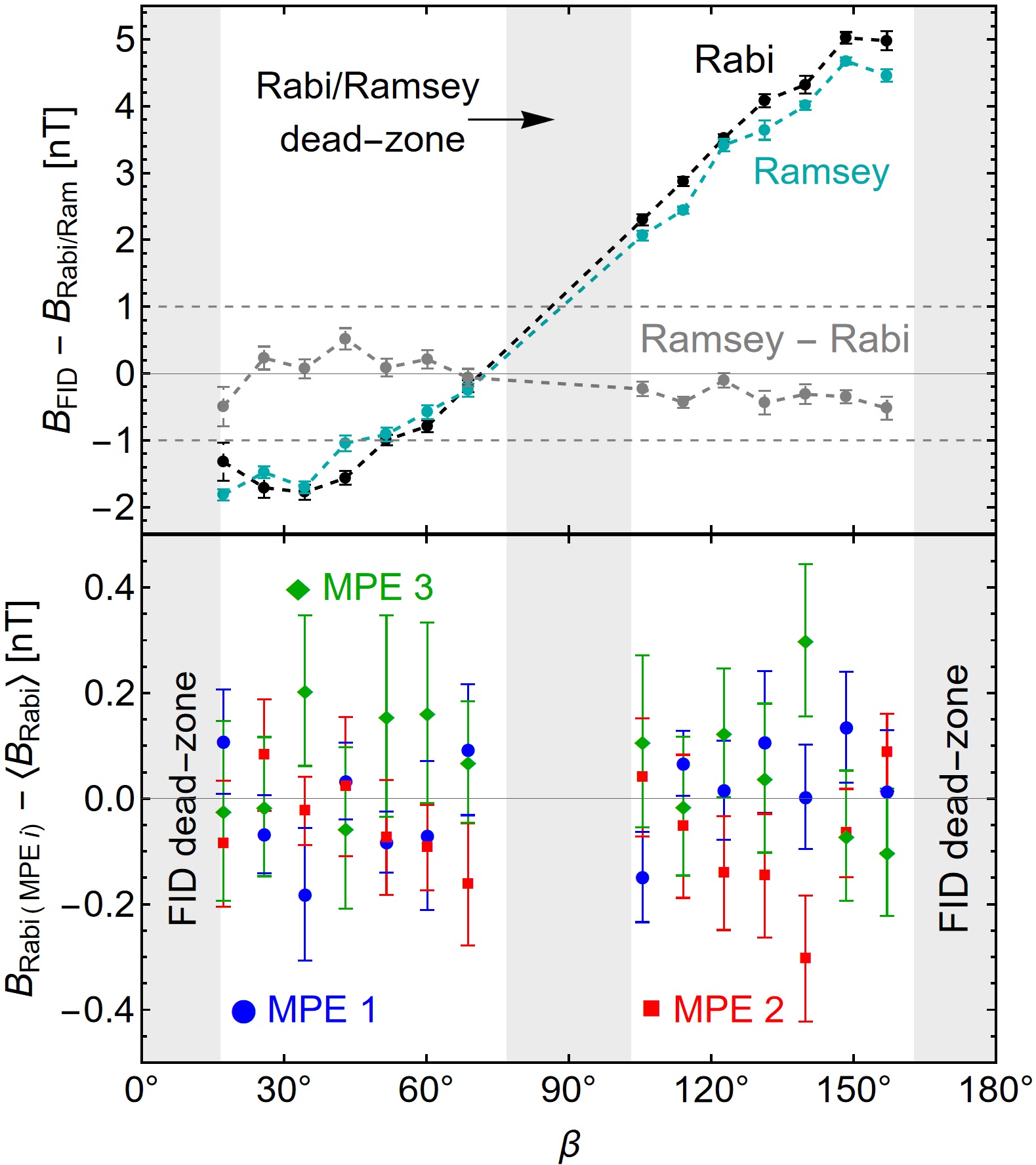}
\end{center}
\caption{Comparison of FID, Rabi, and Ramsey scalar measurements over different magnetic field directions. Error bars show 68\% confidence intervals. Top: differences between FID measurements with Ramsey (cyan) and Rabi (black) scalar measurements, and differences between Ramsey and Rabi scalar measurements (gray). Bottom: fifferences between Rabi scalar measurements for each MPE with respect to the average scalar across all three MPEs.}
\label{fig:finalHeadingErr}
\end{figure}

Final results comparing FID measurements against the magnetic field strengths obtained with Rabi and Ramsey FS are shown in the top plot of Fig.~\ref{fig:finalHeadingErr}. The FID measurements differ from both the Rabi and Ramsey scalar data by up to 5 nT over the 14 magnetic field orientations. Here, magnetic field strengths obtained with Rabi FS are the average of measurements from both of the $(\sigma^-,\pi^-,\sigma^+)$ and $(\sigma^-,\pi^+,\sigma^+)$ configurations displayed in Fig.~\ref{fig:rabMagDifferentTrans}. 

Despite the differing systematic errors predicted from off-resonant microwave driving with various MPEs, the Rabi FS measurements across unique MPEs are consistent to within $\pm 0.3$ nT as shown in the bottom plot of Fig.~\ref{fig:finalHeadingErr}. Proper modeling of atom-microwave coupling is crucial for achieving this level of consistency, as demonstrated in Appendix~\ref{sec:RamseyRabiFSSimulation}. It shows that fitting Rabi measurements using the two-level formalism (see Eq.~\eqref{eq:2lvl}) instead of the $\delta \lambda_m^{m^{\prime}}$ model (see Eq.~\eqref{eq:rabiLambda}) can lead to 4-nT errors due to off-resonant driving.

From theoretical simulations in Appendix~\ref{sec:RamseyRabiFSSimulation}, accounting for $\nu_{\mu \text{w}}$ dependence of the MPE parameters, spin-exchange (SE) frequency shifts~\cite{micalizio2006spin,appelt1998theory}, as well as lineshape distortions from atomic collisions, we estimate scalar errors to be contained within 0.4 nT for Ramsey FS and 0.7 nT for Rabi FS. The larger systematic errors predicted for Rabi FS are due to sensitivity to MPE $\nu_{\mu \text{w}}$-dependence. These errors could be mitigated by improving the flatness of the microwave cavity mode in the region of operation or by calibrating the $\nu_{\mu \text{w}}$ dependence with MPE calibrations. These simulations also show that a large portion of errors ($<0.4$ nT) arise due to frequency shifts from SE collisions. The fact that the Rabi measurements across different MPEs and Ramsey measurements all agree to within 0.6 nT in Fig.~\ref{fig:finalHeadingErr} is confirmation towards these error estimates.  Even so, some experimental discrepancy between Rabi and Ramsey scalar measurements could be due to experimental drift, as these measurements were taken on different days. 

FID heading errors predicted from optical pumping simulations detailed in Appendix~\ref{sec:FIDHeadingErrorSim} correspond well to those in Fig.~\ref{fig:finalHeadingErr}, except for an overall positive offset. These simulations, which utilize our experimental parameters and account for both ground and excited states during optical pumping modeling, predict a heading error of nearly $-0.4$ nT at $\beta=90^{\circ}$, in contrast to the $+1$-nT heading error predicted in the top plot of Fig.~\ref{fig:finalHeadingErr}. It should be noted that a nonzero heading error at $\beta=90^{\circ}$ is expected from vector light shift effects of the 400-mW pumping beam, which perturbs the effective magnetic field direction during optical pumping. This offset discrepancy remains unresolved, but might be linked to uncertainty in our pump beam parameters, as detailed in Appendix~\ref{sec:FIDHeadingErrorSim}, and to effects not accounted for in the heading error simulations, such as spatial inhomogeneity of the pump optical field caused by absorption and reflections from the uncoated cell windows.

\section{Conclusion}
This work demonstrates how tailored atom-microwave interrogation through Rabi and Ramsey FS reduces OPM heading error in geomagnetic fields to the subnanotesla regime, even in the challenging domains of high buffer gas pressure, utilized in MEMS vapor cells, and regimes of weak optical pumping. Based on these findings, we anticipate that the HFS methods presented here will be useful for benchmarking the accuracy of conventional OPMs, such as those using FID detection, and evaluating errors from off-resonant driving in HFS magnetometers. Demonstration of accurate modeling of atom-microwave coupling also establishes a solid foundation to apply Rabi oscillations towards accurate vector magnetometry~\cite{kiehlVector}.

In addition to accuracy, this work also explored the sensitivity of Rabi FS, reaching down to 80 pT$/\sqrt{\text{Hz}}$ that was limited by technical microwave noise. Through a Cram\'er-Rao lower-bound analysis, we showed that the ultimate sensitivity limit, using measured Rabi oscillation amplitudes and dephasing rates, is 30 pT$/\sqrt{\text{Hz}}$. This limit could be improved further by enhancing the optical pumping efficiency into the $F=1$ manifold. While not as sensitive as FID detection, these results show that Rabi FS, by itself, could be useful for high accuracy applications where extreme sensitivity below 10 pT$/\sqrt{\text{Hz}}$ is unnecessary.

Future efforts aimed at improving the MPE frequency dependence and mitigating SE frequency shifts are expected to further minimize systematic errors in Rabi and Ramsey FS, potentially bringing them down to the 100-pT level. Moreover, implementing the Rabi techniques discussed in this work to an all-optical setup, through the use of two-photon Raman transitions for inducing hyperfine Rabi oscillations, presents a promising avenue for exploration. This all-optical approach could retain the advantages of microwave interrogation methods while also being inherently suited for miniaturization, lower power usage, and decreasing electromagnetic noise emitted by the sensor itself.

\begin{acknowledgments} We acknowledge helpful conversations with Georg Bison, Michaela Ellmeier, and Juniper Pollock, and technical expertise from Yolanda Duerst and Felix Vietmeyer.  This work was supported by DARPA through ARO Grant No. W911NF-21-1-0127 and No. W911NF-19-1-0330, NSF QLCI Grant-OMA - 2016244, and the Baur-SPIE Chair in Optical Physics and Photonics at JILA.
\end{acknowledgments}

\appendix

\section{Coil system details}
\label{sec:coilSystem}
The coil system consists of three near-orthogonal coil pairs that generate fields along coil directions $(\vec{x}_c,\vec{y}_c,\vec{z}_c)$ given by
\begin{align}
\vec{B}_{x,c}=I_xa_x(1+\epsilon_x)\vec{x}_c\\
\vec{B}_{y,c}=I_ya_y(1+\epsilon_y)\vec{y}_c\\
\vec{B}_{z,c}=I_za_z(1+\epsilon_z)\vec{z}_c.
\end{align}
Here $(I_x,I_y,I_z)$ are the coil currents in each coil pair, $(a_x,a_y,a_z)=(91.6926,91.2159,392.773)$ $\mu$T/A are precalibrated coil coefficients, and $(\epsilon_x,\epsilon_y,\epsilon_z)$ are coil correction terms to be determined. We establish an orthogonal laboratory frame $\mathcal{L}=(x,y,z)$ with respect to the dc coil system with nonorthogonality angles $(\delta \theta_x, \delta \theta_y, \delta \phi_y)$ through
\begin{align}
\begin{split}
\vec{x}_c=&R_y(\pi/2+\delta \theta_x)\hat{z}=\{\text{cos}[\delta \theta_x],0,-\text{sin}[\delta\theta_x]\}\\
\vec{y}_c=&R_z(\pi/2+\delta \phi_y)R_y(\pi/2+\delta \theta_y)\hat{z}
\\=&\text{cos}[\delta \theta_y]\{-\text{sin}[\delta\phi_y],\text{cos}[\delta \phi_y],-\text{tan}[\delta \theta_y]\}\\
\vec{z}_c=&\hat{z}=\{0,0,1\}.
\end{split}
\end{align}

\begin{figure}[!htb]
\centering
\includegraphics[width=.45\textwidth]{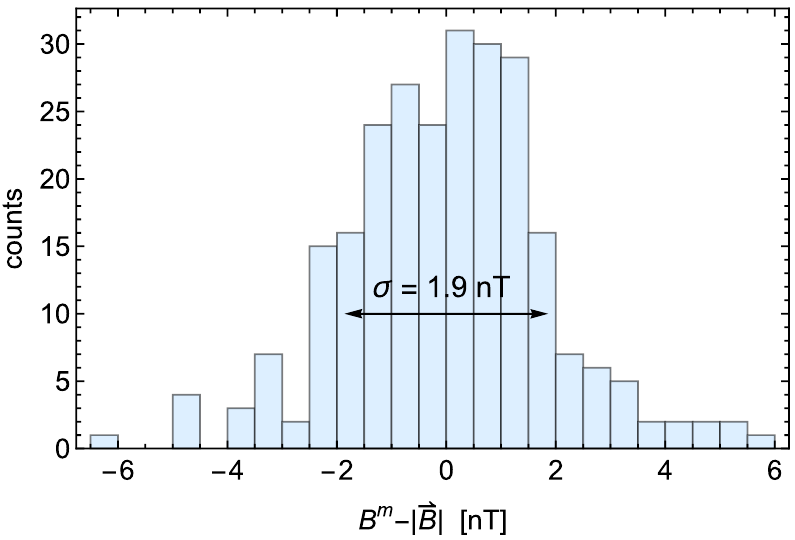}
\caption{Scalar residuals for coil system calibration.}
\label{fig:scalarCalibPlt}
\end{figure}

The total field ($|\vec{B}|$) generated by the coil system and a background field, $\vec{B}_0=(B_{x,o},B_{y,o},B_{z,o})$, is given by
\begin{align}
\begin{split}
\label{eq:coilSystemModel}
|\vec{B}|^2=&\Big(B_{x,o}+\sum_{k=x,y,z}\vec{B}_{k,c}\cdot\hat{x}\Big)^2
\\+&\Big(B_{y,o}+\sum_{k=x,y,z}\vec{B}_{k,c}\cdot\hat{y}\Big)^2\\
+&\Big(B_{z,o}+\sum_{k=x,y,z}\vec{B}_{k,c}\cdot\hat{z}\Big)^2.
\end{split}
\end{align}
In this framework there are nine unknown parameters, namely, three nonorthogonality angles $(\delta \theta_x, \delta \theta_y, \delta \phi_y)=(3.68,-0.91,3.38)$ mrad, three coil corrections $(\epsilon_x, \epsilon_y, \epsilon_z)=(0.82,0.66,2.89)\times10^{-3}$, and three background field components $(B_{x,o},B_{y,o},B_{z,o})=(-77.4,54.4,-70.6)$ nT. These nine parameters are determined by fitting Eq.~\eqref{eq:coilSystemModel} to scalar measurements $B^m\approx 50$ $\mu$T, derived from FID signals across 250 random current configurations. Fig.~\ref{fig:scalarCalibPlt} displays the scalar residuals after performing this calibration. 

\begin{figure}[!htb]
\centering
\includegraphics[width=.45\textwidth]{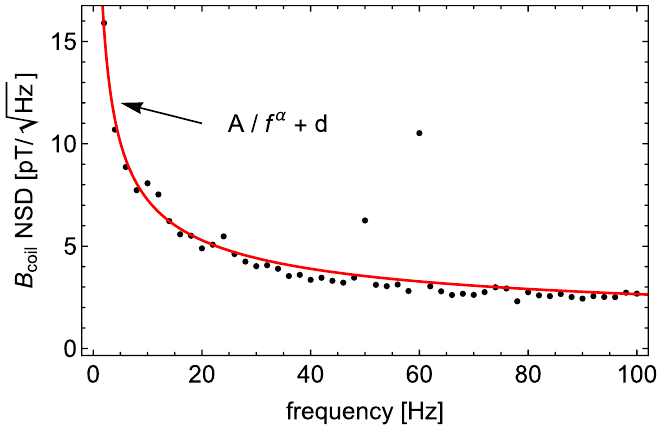}
\caption{The magnetic noise spectral density (NSD) for the $z$-coil pair is derived from the current driver's NSD, with the coil factor $a_z=398.8$ $\mu$T/A. A fit to the NSD shows agreement with 1/$\sqrt{f}$ behavior. The fitted parameters are $A=21.4$, $\alpha=0.502$, and $d=0.538$.}
\label{fig:coilNoise}
\end{figure}

The magnetic noise spectral density of the coil system, calculated based on the noise from the home-built current driver, is illustrated in Fig.~\ref{fig:coilNoise}. Because the coil factors are not identical, the magnetic noise floor, $S_{B_{\text{coil}}}$, depends on the magnetic field direction, and takes the following form in the $x$-$z$ plane as
\begin{equation}
\label{eq:coilNoise}
S_{B_{\text{coil}}}(\beta)=\sqrt{(S_{B_{x,\text{coil}}}\text{sin}(\beta))^2+(S_{B_{z,\text{coil}}}\text{cos}(\beta))^2}.
\end{equation}
The coil noise in Fig.~\ref{fig:rabiSensitivity} is calculated with Eq.~\eqref{eq:coilNoise} using the noise of the coil current driver evaluated at 0.3 Hz, which is the repetition rate for the Rabi measurement sequence shown in Fig.~\ref{fig:RabiScalarSeq} in Appendix~\ref{sec:measSeq}.

\section{Adiabatic optical pumping (AOP)}
\label{sec:aop}
To analyze how the macroscopic atomic spin aligns with the magnetic field during AOP, we consider the Bloch equation
\begin{equation}
\dot{\vec{S}} = \gamma \vec{B} \times \vec{S} -\Gamma_{\text{rel}}\vec{S}+ R\left( \frac{1}{2}s\hat{z} - \vec{S} \right),
\end{equation}
which models the spin dynamics during optical pumping, where $R$ is the pump photon absorption rate, $s\in [ 0,1]$ is the average photon spin along the pumping axis, and $\Gamma_{\text{rel}}$ is the spin-relaxation rate. Without loss of generality, we assume that the magnetic field \(\vec{B} = \{B_x, 0, B_z\}\) is positioned within the $x$-$z$ plane. In steady-state pumping $\dot{\vec{S}}=0$, we can write
\begin{align}
\begin{split}
\label{eq:steadyState}
\gamma (-B_xS_z\hat{y}&+B_xS_y\hat{z}+B_zS_x\hat{y}-B_zS_y\hat{x})\\
&-\Gamma_{\text{rel}}\vec{S}+R(\frac{1}{2}s\hat{z}-\vec{S})=0.
\end{split}
\end{align}
Breaking Eq.~\eqref{eq:steadyState} into each vector component gives
\begin{align}
\begin{split}
&\hat{x}\text{:\quad}-\gamma B_zS_y-(R+\Gamma_{\text{rel}}) S_x=0\\
&\hat{y}\text{:\quad}-\gamma(B_xS_z-B_zS_x)-(R+ \Gamma_{\text{rel}})S_y=0\\
&\hat{z}\text{:\quad}\gamma B_x S_y+R \frac{s}{2}-(R+\Gamma_{\text{rel}})S_z=0.
\end{split}
\end{align}
If $\Gamma_{\text{rel}}=0$ then the $\hat{x}$ equation implies that $S_y=-R S_x/\gamma B_x$. Therefore, if $R \rightarrow 0$ adiabatically then $S_y\rightarrow 0$. In this limit, the $\hat{y}$ equation predicts $B_z/B_x=S_z/S_x$. Consequently, the atomic spin $\vec{S}$ will align with the magnetic field $\vec{B}$, assuming that the pump rate $R$ adiabatically turns off. A similar analysis shows that letting $\Gamma_{\text{rel}}\neq 0$ still allows nearly perfect spin alignment with some spin accumulated in $S_y$ of the order of $|\vec{S}|\Gamma_{\text{rel}}/\gamma |\vec{B}|$. Typically, $\Gamma_{\text{rel}}$ is around a few kilohertz, which causes $S_y/|\vec{S}|\approx 1\%$ for geomagnetic fields. AOP is numerically simulated with a more comprehensive optical pumping model in Appendix~\ref{sec:RamseyRabiFSSimulation} that incorporates all 16 hyperfine states in the ground- and excited-state manifolds.

\section{Measurement sequences}
\label{sec:measSeq}
\begin{figure}[!tbh]
\begin{center}
\includegraphics[width=0.45\textwidth]
{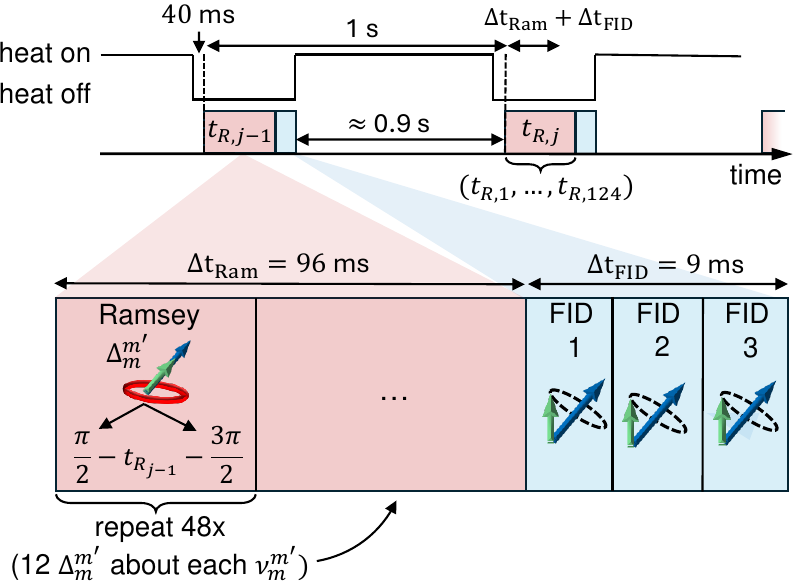}
\end{center}
\caption{Timing diagram for Ramsey FS. Top: Ramsey measurements are segmented into acquisition periods lasting $\Delta t_{\text{Ram}}+\Delta t_{\text{FID}}=105$ ms and repeated every 1 second. A total of 124 acquisition periods are taken that correspond to 124 different Ramsey times $t_{R,j}$ spanning $0.2$ ms to $1.43$ ms. All 124 acquisition periods are repeated 10 times for averaging. Bottom: each acquisition period consists of 48 Ramsey sequences, corresponding to the 12 microwave detunings for each of the four hyperfine transitions $\nu_m^{m^{\prime}}$. At the end of each acquisition period are three FID measurements.}
\label{fig:RamseyScalarSeq}
\end{figure}

Here, we delve deeper into the specifics of the Ramsey, Rabi, and FID measurement sequences. The measurement sequence to perform Ramsey FS [see Fig.~\ref{fig:RamseyScalarSeq}] involves several Ramsey-FID acquisition periods with a repetition rate of 1 second. This includes a 0.9-s dead time between each acquisition period, necessary to ensure that joule heating keeps the cavity temperature stable at $100^{\circ}$C. Each Ramsey-FID acquisition period, lasting $\Delta t_{\text{Ram}}+\Delta t_{\text{FID}}=105$ ms, includes 48 Ramsey sequences evaluated with unique microwave detunings. These 48 detunings correspond to 12 microwave frequencies for each of the four hyperfine transitions, randomly sequenced in time. Following these 48 Ramsey sequences, three FID measurements are carried out. Each acquisition period is repeated 124 times, corresponding to the variation of 124 Ramsey free evolution times within the 48 Ramsey sequences. For averaging, all 124 acquisition periods are repeated 10 times, resulting in a total of 1240 acquisition periods (approximately $34$ min) for each magnetic field measurement. This process, repeated over 14 magnetic field directions, resulted in a total measurement duration of approximately 4.8 h.

\begin{figure}[!tbh]
\begin{center}
\includegraphics[width=0.45\textwidth]
{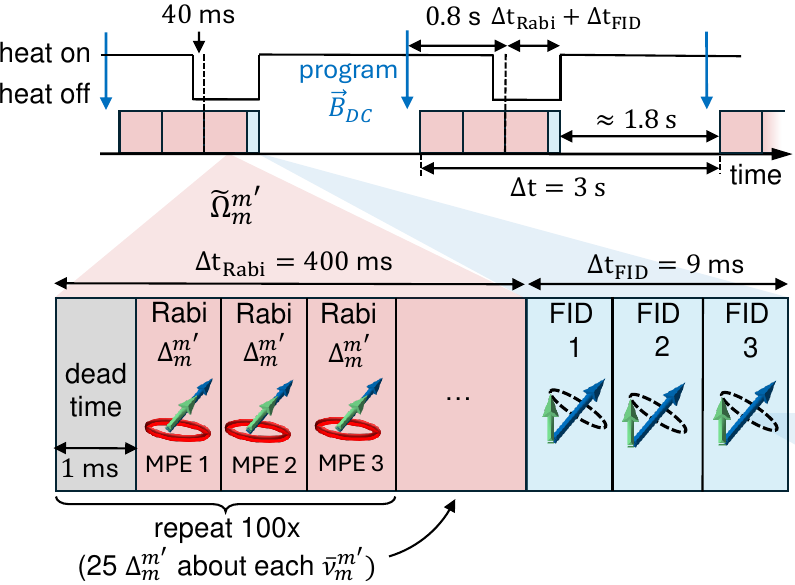}
\end{center}
\caption{Timing diagram for Rabi FS. The Rabi measurement sequence consists of 100 Rabi measurements for each of the 3 MPEs. These 100 Rabi measurements correspond to 25 microwave detunings about each of the four hyperfine transitions shown in Fig.~\ref{fig:Figure1main}(a). As shown in the second row of the timing diagram, Rabi measurements of different MPEs are interlaced at a given microwave frequency. At the end of the acquisition period are three FID measurements.}
\label{fig:RabiScalarSeq}
\end{figure}

For Rabi FS, we utilize the sequence of acquisition periods diagrammed in Fig.~\ref{fig:RabiScalarSeq}. Each Rabi acquisition period, lasting $\Delta t_{\text{Rabi}}=400$ ms, comprises 300 Rabi frequency measurements that correspond to the 25 microwave detunings for each hyperfine transition across the MPE-1, MPE-2, and MPE-3 cavity excitations. Unlike Ramsey FS, all Rabi measurements at a fixed dc magnetic field is contained within the $\Delta t_{\text{Rabi}}$ acquisition period. Similar to the Ramsey measurements, a dead time of roughly 2.6 s is incorporated between each measured Rabi acquisition period to maintain the cavity temperature to near 100 $^{\circ}$C. To allow the microwave components to thermally stabilize, and minimize microwave field drift, we execute the Rabi acquisition period twice before recording the final Rabi acquisition period. The microwave frequencies used in the Rabi measurements are arranged in a random temporal sequence, and different MPE measurements are interlaced to minimize the time any given microwave component is turned off. A 1-ms technical dead time was also incorporated between each set of three MPE measurements, as shown in Fig.~\ref{fig:RabiScalarSeq}. Following the Rabi measurements, three FID measurements are conducted over $\Delta t_{\text{FID}}=9$ ms. For each of the 14 magnetic field orientations, the combined Rabi+FID measurement sequence 
is repeated eight times for averaging.

\section{Rabi oscillation filtering}
\label{sec:RabiFiltering}
The 50-$\mu$s pump power ramp during AOP is not perfectly adiabatic. As a result, a certain level of residual Larmor precession signal persists in our measurements. Larmor precession can cause additional noise and systematic errors during time-domain fitting of the Rabi oscillation. To eliminate the Larmor signal, we utilize a digital equiripple finite impulse response (FIR) filter. 

\begin{figure}[!htb]
\centering
\includegraphics[width=.95\linewidth]{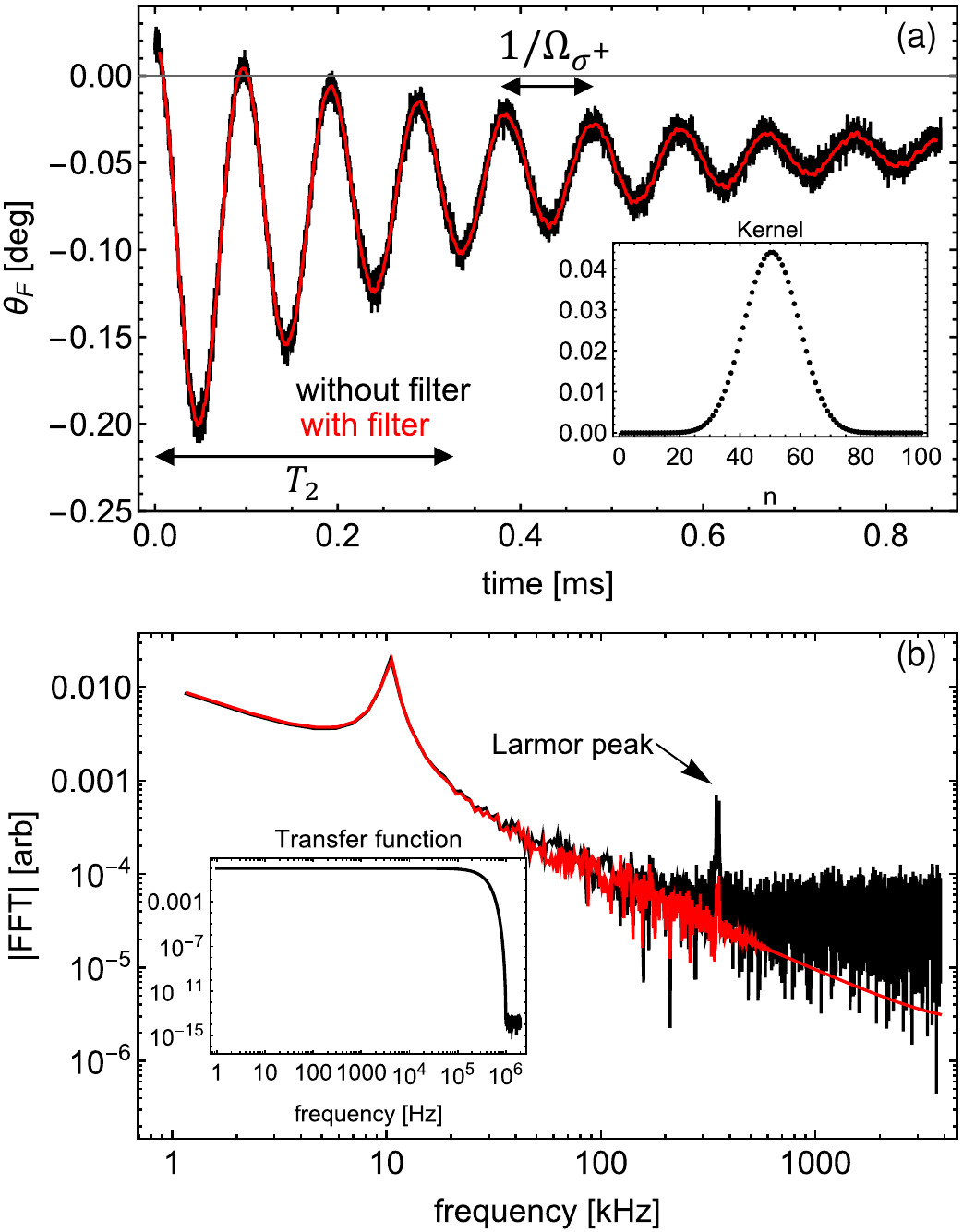}
\caption{Finite impulse response (FIR) filtering of Rabi oscillation measurements evaluated in the (a) time domain and (b) frequency domain.}
\label{fig:rabiFilter}
\end{figure}

The FIR filter's impulse response, along with its application to a Rabi oscillation measurement, is displayed in Fig.~\ref{fig:rabiFilter}. The filter kernel, which is convolved with the measurement data, was generated using the \textsc{mathematica} EquirippleFilterKernel function. The FIR filter is characterized by a passband frequency of 10 kHz, a stopband frequency of 1 MHz, and a filter length of 100.

\section{Magnetic transition dipole moments}
\label{sec:transDipoles}
The magnetic transition dipole moments, found in Eq.~\eqref{eq:Hamiltonian}, for the $\pi$ ($m^{\prime}=m$) and $\sigma^{\pm}$ ($m^{\prime}=m\pm1$) transitions are given by
\begin{align}
\mu_m^{m}=&\mu_B\bra{\overline{2,m}}\mathcal{M}(g_s S_{\pi}+g_i I_{\pi})\mathcal{M}^{\dagger}\ket{\overline{1,m}}\\
\mu_m^{m\pm1}=&\frac{\mu_B\bra{\overline{2,m\pm1}}\mathcal{M}(g_s S_{\pm}+g_i I_{\pm})\mathcal{M}^{\dagger}\ket{\overline{1,m}}}{\sqrt{2}}
\end{align}
where $S_{\pi}=S_z$ and $S_{\pm}=S_x\pm iS_y$ are the electron spin raising and lowering operators with analogous definitions for the nuclear-spin operators $I_{\pi}$ and $I_{\pm}$.  Table~\ref{tab:dipolemoments} tabulates these dipole moments for all of the $5^2S_{1/2}$ hyperfine transitions in $^{87}$Rb.
\begin{table}[ht]
\caption{\label{tab:dipolemoments}
The magnetic transition dipole moments $\mu_m^{m^{\prime}}$ for hyperfine transitions $\ket{1,m}\leftrightarrow \ket{2,m^{\prime}}$. The middle column displays $\mu_m^{m^{\prime}}$ in the limit of $B=0$. The rightmost column displays the relative change in $\mu_m^{m^{\prime}}$ at 50 $\mu$T given by $\delta\mu_m^{m^{\prime}}=[\mu_m^{m^{\prime}}(B=50 \text{ }\mu\text{T})]/[\mu_m^{m^{\prime}}(B=0)]$. }
\begin{center}
\begin{tabular}{|c||c|c|}
\hline
\textrm{Transition} &
\textrm{$\mu_m^{m^{\prime}}$$(B=0)$}&
\textrm{$\delta\mu_m^{m^{\prime}}-1$ $(\%)$}\\
\hline
$\ket{1,1}\leftrightarrow \ket{2,2}$&$-\sqrt{\frac{3}{8}}(g_s-g_i)\mu_B$&$0.42\times10^{-2}$\\
$\ket{1,1}\leftrightarrow \ket{2,1}$&$\frac{\sqrt{3}}{4}(g_s-g_i)\mu_B$&$-1.22\times10^{-2}$\\
$\ket{1,1}\leftrightarrow \ket{2,0}$&$\frac{1}{4}(g_s-g_i)\mu_B$&$-2.66\times10^{-2}$\\
$\ket{1,0}\leftrightarrow \ket{2,1}$&$-\frac{\sqrt{3}}{4}(g_s-g_i)\mu_B$&$1.44\times10^{-2}$\\
$\ket{1,0}\leftrightarrow \ket{2,0}$&$\frac{1}{2}(g_s-g_i)\mu_B$&$0.10\times10^{-2}$\\
$\ket{1,0}\leftrightarrow \ket{2,-1}$&$\frac{\sqrt{3}}{4}(g_s-g_i)\mu_B$&$-1.63\times10^{-2}$\\
$\ket{1,-1}\leftrightarrow \ket{2,0}$&$-\frac{1}{4}(g_s-g_i)\mu_B$&$2.47\times10^{-2}$\\
$\ket{1,-1}\leftrightarrow \ket{2,-1}$&$\frac{\sqrt{3}}{4}(g_s-g_i)\mu_B$&$0.93\times10^{-2}$\\
$\ket{1,-1}\leftrightarrow \ket{2,-2}$&$\sqrt{\frac{3}{8}}(g_s-g_i)\mu_B$&$0.61\times10^{-2}$\\
\hline
\end{tabular}
\end{center}
\end{table}

\begin{table}[ht]
\caption[The operator $\mathcal{M}$ that transforms the hyperfine basis $\ket{F,m_F}$ into the basis $\ket{\overline{F,m_F}}$.]{\label{transformM}%
The operator $\mathcal{M}$ that transforms the hyperfine basis $\ket{F,m_F}$ into the basis $\ket{\overline{F,m_F}}$, which diagonalizes the hyperfine and Zeeman part of the Hamiltonian defined by Eq.~\eqref{eq:Hamiltonian}.}

\begin{center}
\begin{tabular}{c | c c c c c c c c}
\textrm{\quad} &
\textrm{\scriptsize$ \ket{1,1}$} &\textrm{\scriptsize$ \ket{1,0}$} &\textrm{\scriptsize$ \ket{1,-1}$} &\textrm{\scriptsize$ \ket{2,2}$} &\textrm{\scriptsize$ \ket{2,1}$} &\textrm{\scriptsize$ \ket{2,0}$} &\textrm{\scriptsize$ \ket{2,-1}$} &\textrm{\scriptsize$ \ket{2,-2}$}\\[1mm]
\hline
\\ [-.3em]
\textrm{\scriptsize$ \ket{\overline{1,1}}$} &$\mathcal{M}_{11}$ &0 &0 &0 &$-\mathcal{M}_{15}$ &0&0&0\\[2mm]
\textrm{\scriptsize$ \ket{\overline{1,0}}$} &0 &$\mathcal{M}_{22}$ &0 &0 &0 &$-\mathcal{M}_{26}$&0&0\\[2mm]
\textrm{\scriptsize$ \ket{\overline{1,-1}}$} &0 &0 &$\mathcal{M}_{33}$ &0 &0 &0&$-\mathcal{M}_{37}$&0\\[2mm]
\textrm{\scriptsize$ \ket{\overline{2,2}}$} &0 &0 &0 &1 &0 &0&0&0\\[2mm]
\textrm{\scriptsize$ \ket{\overline{2,1}}$} &$\mathcal{M}_{15}$ &0 &0 &0 &$\mathcal{M}_{55}$ &0&0&0\\[2mm]
\textrm{\scriptsize$ \ket{\overline{2,0}}$} &0 &$\mathcal{M}_{26}$ &0 &0 &0 &$\mathcal{M}_{66}$&0&0\\[2mm]
\textrm{\scriptsize$ \ket{\overline{2,-1}}$} &0 &0 &$\mathcal{M}_{37}$ &0 &0 &0&$\mathcal{M}_{77}$&0\\[2mm]
\textrm{\scriptsize$ \ket{\overline{2,-2}}$} &0 &0 &0 &0 &0 &0&0&$1$\\[2mm]
\end{tabular}
\end{center}
\end{table}

Calculation of these dipole moments requires the operator $\mathcal{M}$, which transforms the total atomic spin basis $\ket{F,m_F}$ into the basis $\overline{\ket{F,m_F}}$. This new basis diagonalizes the hyperfine and Zeeman structure of $H$ defined in Eq.~\eqref{eq:Hamiltonian}. While the effect of the pressure shift $\nu_{\text{bg}}\approx 88$ kHz  on $\mathcal{M}$ is negligible, the magnetic field-strength dependence of $\mathcal{M}$ is also small, but not negligible. The third column of Table~\ref{tab:dipolemoments} diplays the relative change of $\mu_m^{m^{\prime}}$ at $B=50$ $\mu$T from $B=0$. Table \ref{transformM} shows the magnetic field-strength dependence of $\mathcal{M}$ that is calculated by fitting polynomial terms to the $\mathcal{M}$-matrix elements for  $B\in\{0,1 \}$ mT. The explicit functions for this magnetic field dependence are given by
\begin{align}
\begin{split}
 \mathcal{M}_{11}=\mathcal{M}_{55}=&1-1.57774B^2+6.47256B^3\\
 &-6.35276B^4
 \end{split}
\end{align}
\begin{align}
\begin{split}
 \mathcal{M}_{22}=\mathcal{M}_{66}=&1-2.10366B^2-0.0000379091B^3\\
 &+24.3589B^4
\end{split}
\end{align}
\begin{align}
\begin{split}
 \mathcal{M}_{33}=\mathcal{M}_{77}=&1-1.57774B^2-6.47264B^3\\&-6.04991B^4
 \end{split}
\end{align}
\begin{align}
\begin{split}
 \mathcal{M}_{15}=&1.77637B-3.64364B^2-0.933969B^3\\
 &+36.1278B^4
  \end{split}
\end{align}
 \begin{align}
\begin{split}
 \mathcal{M}_{26}=&2.05117B+1.16664\times10^{-7}B^2\\
 &-12.9456B^3+0.35176B^4
   \end{split}
\end{align}
 \begin{align}
\begin{split}
 \mathcal{M}_{37}=&1.77637B+3.64364B^2-0.933969B^3\\
 &-36.6891B^4
 \end{split}
\end{align}
where $B$ is in units of Tesla.

\section{Eigenvalue selection for Rabi FS}
\label{sec:EigenvalueSelect}
\begin{figure}[!htbp]
\centering
\includegraphics[width=1\linewidth]{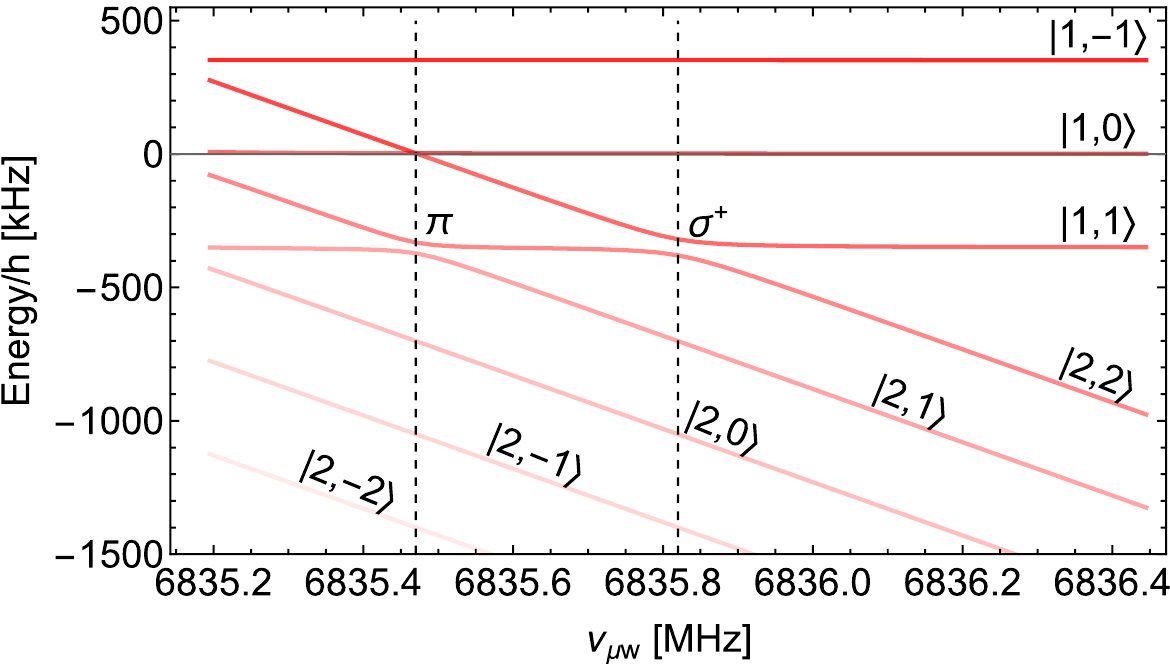}
\caption{Energy eigenvalues ($\lambda_i$) plotted in a frame rotating at $\nu_{\mu\text{w}}$ for a magnetic field strength $B=50$ $\mu$T and microwave parameters $\mathcal{B}_x=3.5$ $\mu$T, $\mathcal{B}_x=5.9$ $\mu$T, $\mathcal{B}_z=0.1$ $\mu$T, $\phi_x=2.6$ rad, $\phi_y=4.0$ rad. Anticrossings occur at the hyperfine resonances (dashed lines).}
\label{fig:rabiEnergyLevels}
\end{figure}

Here, we outline an algorithm to find the correct pair of eigenvalues $(\lambda_j,\lambda_i)$ of the atom-microwave Hamiltonian $H$ such that $\tilde{\Omega}_m^{m^{\prime}}=(\lambda_j-\lambda_i)/h$ for any hyperfine transition. First, we diagonalize the atom-microwave Hamiltonian $H$, defined in Eq.~\eqref{eq:Hamiltonian}, to obtain $H_d$ through the transformation $H_d = D^* H D^{T}$, where the rows of matrix $D$ consist of the eigenvectors of $H$. Then, the diagonal elements, $\{\lambda_1,...,\lambda_8\}$, of $H_d$ are the eight eigenvalues of $H$. For illustration, these eigenvalues are plotted in Fig.~\ref{fig:rabiEnergyLevels} for specific microwave parameters.

Next, let $\rho_m^{m^{\prime}}$ be the density matrix with all of the atomic population in the $\ket{2,m^{\prime}}$ state and $\mathcal{F}=  F_{z,b}-F_{z,a}$
be the difference between the z-component operators of the total atomic spin for the $F = 2$ ($F_{z,b}$) and $F = 1$ ($F_{z,a}$) manifolds. The expectation value of operator $\mathcal{F}$ is proportional to the Faraday rotation signal (see Eq.~\eqref{eq:faradayRotationOperator}). Then, the amplitude of population dynamics corresponding to the energy difference of the eigenvalue pair ($\lambda_j,\lambda_i$) is given by
\begin{equation}
\label{eq:rabiEigenAmp}
a_{i,j}=|(D^*\rho_m^{m^{\prime}}D^{T})_{ij} (D^*\mathcal{F}D^{T})_{ji}|.
\end{equation}
Because the density matrix ($\rho_m^{m^{\prime}}$) is defined to encompass the atomic population solely within one of the sublevels of the targeted hyperfine transition, the indices yielding the highest $a_{i,j}$ match the appropriate pair of eigenvalues, from the set $\{\lambda_1,...,\lambda_8\}$, such that $\tilde{\Omega}_m^{m^{\prime}}=(\lambda_j-\lambda_i)/h$.

\section{Characterization of the microwave polarization ellipses (MPEs)}
\label{sec:MPECharacterization}
To characterize the three MPEs employed in Rabi FS, we fit the five MPE parameters ($\mathcal{B}_x,\mathcal{B}_y,\mathcal{B}_z,\phi_x,\phi_y$) from generalized Rabi frequency measurements at each of the 14 magnetic field directions with the eigenvalue value model $\delta \lambda_m^{m^{\prime}}$ defined in Eq.~\eqref{eq:rabiLambda}.

\begin{figure}[tbh]\centering
\includegraphics[width=.40\textwidth]{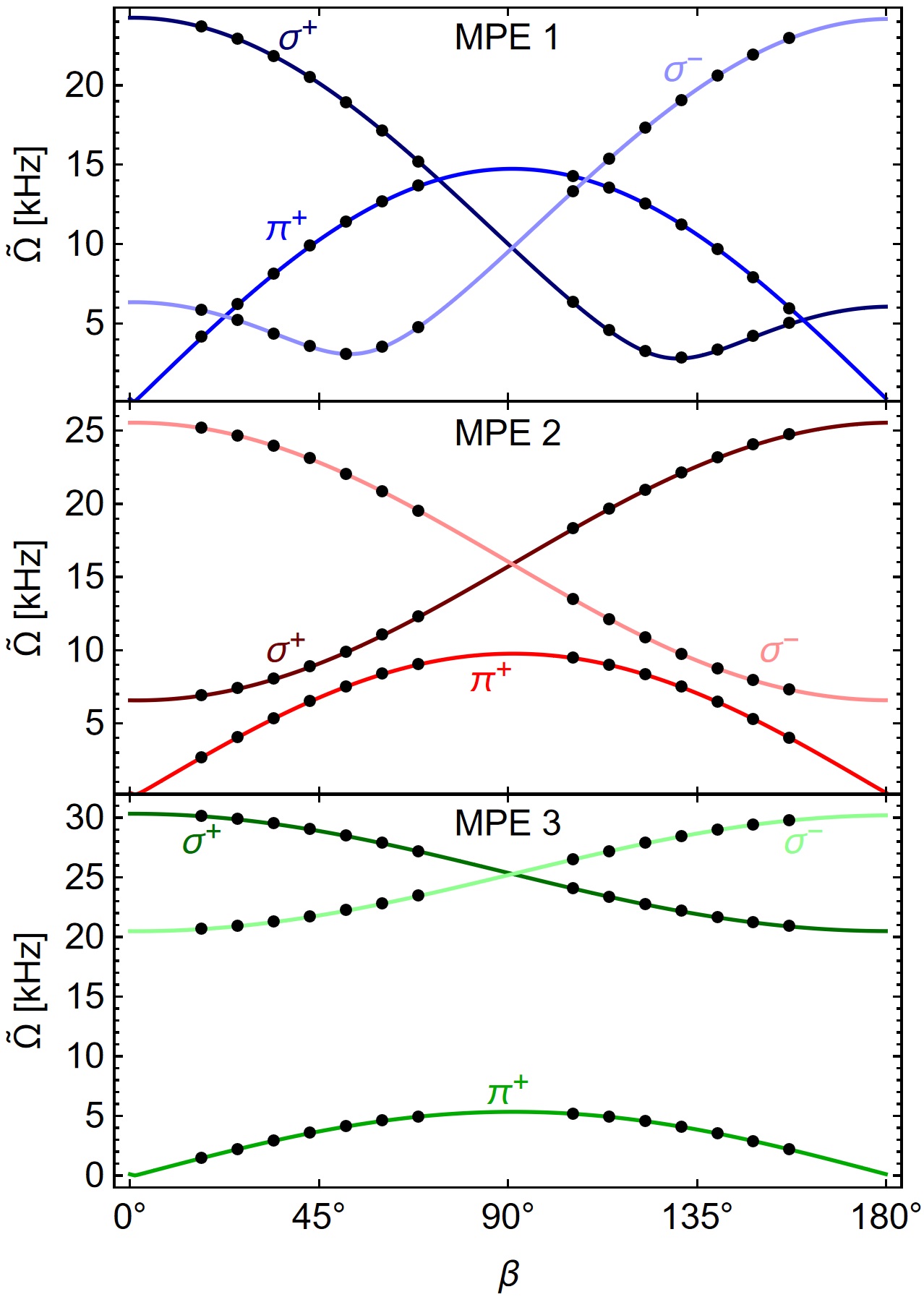}
\caption{Polar angle ($\beta$) dependence of the Rabi frequency measurements for the $\sigma^+,\pi^{+}$, and $\sigma^-$ hyperfine transitions. Solid lines show fits, using Eq.~\eqref{eq:rabiLambda}, to calibrate the MPE parameters.}
\label{fig:rabiCalibrateMPEs}
\end{figure}

We make 12 independent MPE fits corresponding to the Rabi measurements driven at the four microwave frequencies $\nu_{\mu\text{w}}=\overline{\nu}_m^{m^{\prime}}$, and the three cavity excitations. We fix the pressure shift to $\nu_{\text{bg}}=88.02$ kHz during these fits, which is consistent with the measurements in Fig.~\ref{fig:rabMagDifferentTrans}(b,d). The $\delta \lambda_m^{m^{\prime}}$ model with the calibrated MPE parameters, along with the measured generalized Rabi frequencies ($\tilde{\Omega}_m^{m^{\prime}}$), are shown in Fig.~\ref{fig:rabiCalibrateMPEs}. Within these fits, the Rabi measurements are weighted in terms of the generalized Rabi frequency fitting error, and the magnetic field strength ($B$) is known from FID measurements. FID systematic errors are not a concern for these calibrations because $\tilde{\Omega}_m^{m^{\prime}}$ depends on $\Delta_m^{m^{\prime}}$ to second-order near the transition resonance (see Eq.~\eqref{eq:2lvl}).

The calibrated MPE parameters for each of the three MPEs at each microwave frequency $\overline{\nu}_m^{m^{\prime}}$ are tabulated in Table~\ref{tab:PEfreqDep} and plotted in Fig.~\ref{fig:calibratedMPEs}. We utilize these MPE parameters in Appendix~\ref{sec:RamseyRabiFSSimulation} to study a potential systematic error in Rabi FS due to $\nu_{\mu\text{w}}$ dependence of the MPE parameters arising from the microwave cavity mode linewidth ($\Gamma=110$ MHz). For comparison, the microwave frequencies of our 
Rabi measurements span nearly 2 MHz [see Fig.~\ref{fig:rabiFreqSpecn2}]. Other factors, such as standing waves in the SMA cables and frequency-dependent amplitude and phase shifts from the bandpass filters and microwave amplifiers, may also contribute to this frequency dependence.

\begin{table}[!tbh]
\caption{\label{tab:PEfreqDep}%
MPE 1 (rows 1-4), MPE 2 (rows 5-8), and MPE 3 (rows 9-12) parameters calibrated at microwave frequencies $\overline{\nu}_{m}^{m^{\prime}}$, which characterize the microwave field used during Rabi FS.}
\begin{center}
\begin{tabular}{c | c c c c c}
$\nu_{\mu\text{w}}$ (MHz) &
\textrm{$ \mathcal{B}_x$ ($\mu$T)} &\textrm{$ \mathcal{B}_y$ ($\mu$T)} &\textrm{$ \mathcal{B}_z$ ($\mu$T)} &\textrm{$ \phi_x$ (rad)} &\textrm{$ \phi_y$ (rad)}\\[1mm]
\hline
\\ [-.3em]
\textrm{\scriptsize$\overline{\nu}_{-1}^{-2}=6833.7203$} &1.2177 &0.7972 &0.0189 &2.727 &0.8424\\[2mm]
\textrm{\scriptsize$\overline{\nu}_{-1}^{-1}=6834.0701$} &1.2156 &0.7644 &0.0200 &2.7801 &0.7944\\[2mm]
\textrm{\scriptsize$\overline{\nu}_{1}^{1}=6835.472$} &1.2136 &0.7661 &0.0214 &2.768 &0.7415\\[2mm]
\textrm{\scriptsize$\overline{\nu}_{1}^{2}=6835.8218$} &1.2104 &0.8095 &0.0226 &2.571 &0.6976\\[2mm]
\hline
\\ [-.3em]
\textrm{\scriptsize$\overline{\nu}_{-1}^{-2}=6833.7203$} &0.8043 &1.3088 &0.0304 &4.095 &5.858\\[2mm]
\textrm{\scriptsize$\overline{\nu}_{-1}^{-1}=6834.0701$} &0.8078 &1.3602 &0.0243 &3.894 &5.541\\[2mm]
\textrm{\scriptsize$\overline{\nu}_{1}^{1}=6835.472$} &0.8039 &1.3327 &0.0216 &3.768 &5.849\\[2mm]
\textrm{\scriptsize$\overline{\nu}_{1}^{2}=6835.8218$} &0.7944 &1.3155 &0.0264 &3.949 &5.370\\[2mm]
\hline
\\ [-.3em]
\textrm{\scriptsize$\overline{\nu}_{-1}^{-2}=6833.7203$} &0.4118 &2.0851 &0.0120 &2.899 &1.0936\\[2mm]
\textrm{\scriptsize$\overline{\nu}_{-1}^{-1}=6834.0701$} &0.4461 &2.1708 &0.0067 &3.155 &0.9992\\[2mm]
\textrm{\scriptsize$\overline{\nu}_{1}^{1}=6835.472$} &0.4405 &1.9872 &0.0093 &2.912 &1.0956\\[2mm]
\textrm{\scriptsize$\overline{\nu}_{1}^{2}=6835.8218$} &0.4345 &2.0876 &0.0032 &2.970 &1.0365\\[2mm]
\end{tabular}
\end{center}
\end{table}

\section{Characterization of polarimeter noise}
\label{sec:polarimeterNoise}

\begin{figure}[!htb]
\centering
\includegraphics[width=0.45\textwidth]{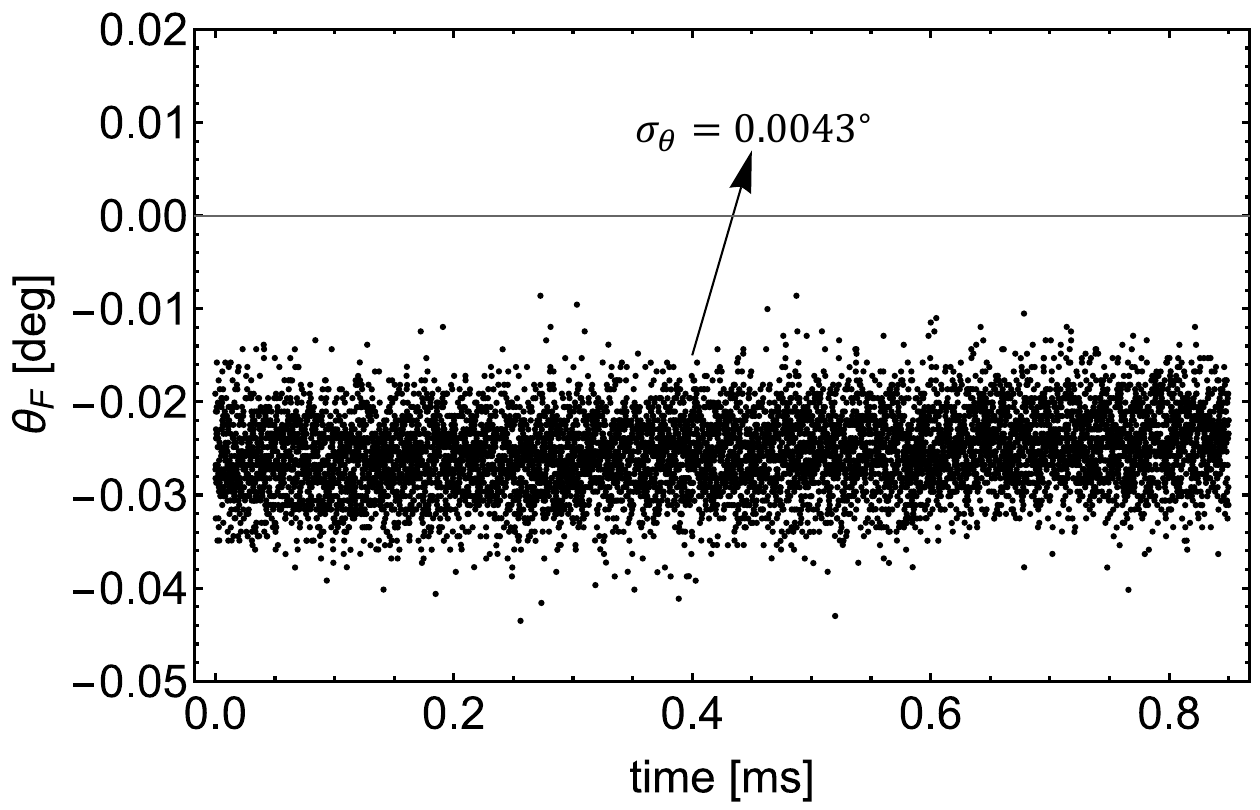}
\caption{Faraday rotation noise measurement.}
\label{fig:shotNoise}
\end{figure}

Faraday rotation is detected with two photodiodes that individually measure the horizontal and vertical components of the probe beam polarization. The Faraday rotation angle ($\theta_F$) is calculated from these photodiode signals, $P_1$ and $P_2$, as
\begin{equation}
\label{eq:rotFromPhotodiode}
    \theta_F=\frac{1}{2}\text{arcsin}\Big(\frac{P_2-P_1}{P_1+P_2}\Big).
\end{equation}
The detector noise is characterized by the standard deviation $\sigma_{\theta}=0.0043^{\circ}$ measured over the 0.85-ms duration of a typical Rabi oscillation [see Fig.~\ref{fig:shotNoise}]. This detector noise is near the optical shot-noise limit characterized by the sensitivity~\cite{lucivero2014shot} 
\begin{equation}
\label{eq:shotnoise}
S_{\theta_F,\text{shot}}=\frac{1}{2}\frac{1}{\sqrt{\dot{N}_{\text{ph}}}}
\end{equation}
where $\dot{N}_{\text{ph}}=P/(hc/\lambda)$ is the number of photons in the probe beam per second and $P=1$ mW is the probe optical power. Equation~\eqref{eq:shotnoise} is derived from Eq.~\eqref{eq:rotFromPhotodiode} by making the small-angle approximation and writing
\begin{equation}
\theta_F\approx \frac{1}{2}\frac{\dot{N}_{\text{ph},2}-\dot{N}_{\text{ph},1}}{\dot{N}_{\text{ph},1}+\dot{N}_{\text{ph},2}}
\end{equation}
and assuming that each photodetector collects $\dot{N}_{\text{ph},i}=\dot{N}_{\text{ph}}/2$ photons per unit time with shot-noise uncertainty $\sqrt{\dot{N}_{\text{ph}}/2}$.
At 1-mW probe power, Eq.~\eqref{eq:shotnoise} predicts a standard deviation $\sigma_{\text{shot}}=S_{\theta_{F},\text{shot}}\sqrt{f_s/2}=0.0033^{\circ}$ in the Faraday rotation signal, where $f_s=10$ MHz is the detector sampling rate.

\section{Theoretical model for Rabi, Ramsey, and FID atomic spin dynamics}
\label{sec:ScalarTheoritcalSim}

Here, we outline the theoretical model for atomic spin dynamics during Rabi, Ramsey, and FID measurements. We utilize this model to generate simulated measurements, which are then used to test the Ramsey and Rabi FS fitting, and to characterize the magnitude of potential systematic errors, as discussed in Appendix~\ref{sec:RamseyRabiFSSimulation}. Additionally, we model optical pumping, detailed in Appendix~\ref{sec:fullopticalPumpModel}, to estimate the FID heading error based on our specific pumping beam and vapor cell parameters, with the results presented in Appendix~\ref{sec:FIDHeadingErrorSim}.

Given the initial $8\times8$ density matrix $\rho_0$ following optical pumping, which describes the 5$^2$S$_{1/2}$ ground-state manifold 
in the $\ket{F,m_F}$ basis, the time evolution of $\rho_0$, including multiple sources of collisional relaxation, is given by (see Refs.~\cite{budker2013optical,allred2002high,jau2005new})
\begin{align}
\label{eq:timeEvo}
\begin{split}
    \dot{\rho}=&\frac{[H^{(\alpha,\beta)},\rho]}{i \hbar}+\frac{[\delta \mathcal{E}_{\text{se}},\rho]}{i \hbar}+\dot{\rho}_{\text{se}}+\dot{\rho}_{\text{sd}}+\dot{\rho}_{\text{c}}+\dot{\rho}_{\text{D}}\\=&\frac{[H,\rho]}{i \hbar}+\frac{[n_{\text{Rb}}\lambda_{\text{se}}v_r \langle \mathbf{S} \rangle\cdot \mathbf{S},\rho]}{i}
    \\&+\Gamma_{\text{se}}\big(\phi(1+4\langle \textbf{S} \rangle\cdot \textbf{S})-\rho\big)+\Gamma_{\text{sd}}\big(\phi-\rho\big)
    \\&-\frac{\eta_I^2 [I]^2}{8}\Gamma_{\text{C}} \rho^{(\text{m})}+  \Gamma_D(\rho^e-\rho)
\end{split}
\end{align}
where
\begin{equation}
\label{eq:rotatedHamiltonian}
H^{(\alpha,\beta)}=e^{-iF_z\alpha}e^{-iF_y\beta}He^{iF_y\beta}e^{iF_z\alpha}
\end{equation}
is the rotated Hamiltonian to account for an arbitrary magnetic field direction $(\alpha,\beta)$, with $H$ being the atom-microwave Hamiltonian defined in Eq.~\eqref{eq:Hamiltonian}. To simulate free evolution with the microwave field turned off, as occurs during FID measurements and Ramsey interferometry, the spherical microwave components $(\mathcal{B}_k^{(\alpha,\beta)})$ in Eq.~\eqref{eq:Hamiltonian} are set to zero. The term $\phi=\rho/4+\textbf{S}\cdot \rho\textbf{S}$ in Eq.~\eqref{eq:timeEvo} is known as the nuclear part of the density matrix, for which Tr$[\phi \textbf{S}]=0$ and Tr$[\phi]=1$~\cite{appelt1998theory}. The operators $F_x$, $F_y$, and $F_z$ in Eq.~\eqref{eq:rotatedHamiltonian} denote the Cartesian components of the total atomic spin operator $\mathbf{F}=\mathbf{S}+\mathbf{I}$. 

\begin{table*}[t]
\caption[Collision cross sections and rates for simulating optical pumping]{\label{tab:collisionRatesHeading}
The collision mechanisms, cross sections $\sigma_i$, and the calculated collision rates for a vapor cell with volume $3\times3\times2$ mm$^3$, vapor temperature $\mathcal{T}_v=100^{\circ}$C, buffer gas pressure $\text{P}_{\text{N}_2}=180$ Torr (24 kPa), and diffusion constant $D_0= 0.216$ cm$^2$s$^{-1}$ for Rb-N$_2$ buffer gas collisions scaled to our vapor temperature~\cite{pouliot2021accurate}. The tabulated mechanisms are quenching, optical dephasing (OD), spin-destruction (SD), spin-exchange (SE), wall collisions (WC), and Carver relaxation.}
\begin{center}
\begin{tabular}{|c|c|c|}
\hline
\textrm{Collision type} &
\textrm{cross-section [$10^{-18}$ m$^2$]}&
\textrm{collision rate}
\\
\hline
5P$_{1/2}$ quench (Rb-N$_2$)&$\sigma_q=0.58$~\cite{hrycyshyn1970inelastic,seltzer2008developments} &$\Gamma_q=n_{\text{N}_2}\sigma_q v_r=1.65\times 10^9$ s$^{-1}$\\
5P$_{1/2}$ OD (Rb-N$_2$)&- &$\Gamma_o=2\pi\cdot 5.6$ GHz (measured)\\
5P$_{1/2}$ SD (Rb-N$_2$) &$\sigma_p=0.64$~\cite{happer2010optically} &$\Gamma_p=n_{\text{N}_2}\sigma_p v_r=1.82\times 10^9$ s$^{-1}$\\
5S$_{1/2}$ SD (Rb-N$_2$) & $\sigma_{sd}=1.44\times 10^{-8}$~\cite{wagshul1989optical}&$\Gamma_{sd}=n_{\text{N}_2}\sigma_{sd} v_r=41$ s$^{-1}$\\
5S$_{1/2}$ SD (Rb-Rb) & $\sigma_{sd} =1.77\times 10^{-3}$~\cite{wagshul1989optical}&$\Gamma_{sd}=n_{\text{Rb}}\sigma_{sd} v_r=3.6$ s$^{-1}$\\
5S$_{1/2}$ SE (Rb-Rb) & $\sigma_{se}=1.9$~\cite{gibbs1967spin} ($\lambda_{\text{se}}=0.69$~\cite{micalizio2006spin})&$\Gamma_{se}=n_{\text{Rb}}\sigma_{se} v_r=3.89\times 10^3$ s$^{-1}$\\
5S$_{1/2}$ WC& $D_0P_0=0.016 \text{ m}^2\text{Torr}$ s$^{-1}$~\cite{pouliot2021accurate}&$\Gamma_{\text{D}}=\frac{D_0 \text{P}_0\pi^2}{\text{P}_{\text{N}_2}(l_x^2+l_y^2+l_z^2)}=0.45\times 10^3$ s$^{-1}$~\cite{kiehl2023coherence}\\
5S$_{1/2}$ Carver (Rb-N$_2$)& $\Gamma_{\text{C}}/[\text{N}_2]=394 \text{ amg}^{-1} \text{s}^{-1}$~\cite{walter2002magnetic} & $\Gamma_{\text{C}}=69$ s$^{-1}$\\
\hline
\end{tabular}
\end{center}
\end{table*}

The terms $\dot{\rho}_{\text{se}}$ and $\dot{\rho}_{\text{sd}}$ in Equation~\eqref{eq:timeEvo} represent the relaxation processes due to spin-exchange (SE) and spin-damping collisions. The collision rates $\Gamma_{\text{se(sd)}}=n_{\text{a(N$_2$)}}\sigma_{\text{se(sd)}}v_r$ for these processes are defined by the cross sections $\sigma_{\text{se(sd)}}$, the mean relative velocity $v_r$ of the colliding entities, and $n_{a}$,$n_{\text{N}_2}$, representing the atomic densities for alkali and buffer gas collisions, respectively~\cite{allred2002high}. Additionally, the term $\delta \mathcal{E}_{\text{se}}$ of Eq.~\eqref{eq:timeEvo} accounts for frequency shifts from SE collisions~\cite{vanier1989quantum,appelt1998theory} proportional to the cross section $\lambda_{\text{se}}=0.69\times10^{-18}$ m$^2$~\cite{micalizio2006spin}. The term $\dot{\rho}_{\text{c}}$ models pure dephasing of microwave transitions due to buffer gas collisions, where $\rho^{(\text{m})}$ signifies the density matrix with only off-diagonal terms that represent the coherences between the upper and lower hyperfine manifolds. Here, $\Gamma_{\text{C}}$ is the Carver rate, and $\eta_{I}=\mu_I/(2 I \mu_N)$ is the isotope coefficient~\cite{jau2005new,walter2002magnetic}, where $\mu_I$ and $\mu_N$ are the nuclear magnetic moment and the nuclear magneton, respectively. The final term in Eq.~\eqref{eq:timeEvo} models diffusion into the cell walls where alkali spins are completely randomized. Here $\rho^e$ is the equilibrium density matrix
with all populations $\rho^e_{ii}$ equal and $\Gamma_{\text{D}}=D\pi^2/(l_x^2+l_y^2+l_z^2)$ is the fundamental decay mode defined by the vapor cell dimensions $l_x$, $l_y$, and $l_z$~\cite{franzen1959spin}. The diffusion constant $D=D_0 \text{P}_0/\text{P}_{\text{N}_2}$ is attenuated by the buffer gas pressure where $\text{P}_0=1$ atm. For context, Table~\ref{tab:collisionRatesHeading} contains the collision rates, cross sections, and the diffusion constant assuming the vapor cell parameters and the Rb-N$_2$ alkali-buffer gas mixture used in our experiment. 

Because Hamiltonian $H$ in Eq.~\eqref{eq:Hamiltonian} is defined in the RWA, we must place the decoherence operators, e.g. $\Gamma_{\text{se}}\big(\phi(1+4\langle \mathbf{S} \rangle \cdot \mathbf{S} )-\rho \big)$, in the rotating frame and make the RWA accordingly. The transition into the frame rotating at the microwave frequency $\omega_{\mu\text{w}}=2\pi \nu_{\mu\text{w}}$ is achieved with the diagonalized unitary operator $\mathcal{U}(t)$, characterized by elements  $\mathcal{U}_{ii}(t)=1$ [$\mathcal{U}_{ii}(t)=e^{\text{-}i\omega_{\mu\text{w}} t}$] for states in the $F=1$ ($F=2$) manifold. To make the RWA on the decoherence operators, we first put the spin matrices in the rotating frame $\mathbf{S}\rightarrow\mathbf{S}(t)=\mathcal{U}^{\dagger}(t)\mathbf{S} \mathcal{U}(t)$ and then numerically average the decoherence operators with time steps $\Delta t= 1/(\nu_{\mu\text{w}} N_{\text{ave}})$, where $N_{\text{ave}}$ is the number of averages. We cannot average the spin matrices individually since the decoherence operators contain higher-order products like $S_x(t)\rho(t) S_x(t)$ that contain nontrivial cancelation of counter-rotating terms before averaging. Instead, we calculate
\begin{align}
\begin{split}
  \tilde{S}_x(t)\rho(t) S_x(t)  \rightarrow  &\frac{1}{N_{\text{ave}}}\sum_{k=0}^{N_{\text{ave}}-1} S_x(k \Delta t) \rho (t) S_x(k \Delta t)
\end{split}
\end{align}
such that high-frequency terms $\propto e^{\pm i n \omega_{\mu\text{w}}t}$ oscillating at multiple integers of $\nu_{\mu \text{w}}$ are eliminated. Note that, since we are using the time-independent Hamiltonian $H$, $\rho(t)$ in the rotated frame contains no counter-rotating terms and is left fixed at time $t$ during the averaging. Here, we use $N_{\text{ave}}=4$, where the $e^{\pm i n \omega_{\mu\text{w}}t}$ terms in this rotated frame are efficiently averaged away using time steps $\Delta t$.

In this framework, the Faraday rotation signal is given by 
\begin{equation}
\label{eq:faradayRotationOperator}
\theta_F\propto \langle \mathcal{F} \rangle = \langle F_{z,b}-F_{z,a} \rangle,
\end{equation}
where $\langle F_{z,a} \rangle$ and $\langle F_{z,b} \rangle$ denote the expectation values of the z component of the hyperfine spin for the
$F = 1$ and $F = 2$ manifolds, respectively. We note that $\langle \mathcal{F} \rangle = 4\langle S_z \rangle$, where off-diagonal elements of $S_z$  expressed in the $\ket{F,m_F}$ basis average to zero in the rotating-wave approximation.

\section{Optical pumping model with excited and ground hyperfine sublevels}
\label{sec:fullopticalPumpModel}
In order to determine the optically pumped atomic state $\rho_0\equiv\rho_{\text{gg}}$, from which Rabi, Ramsey, and FID signals are simulated using Eq.~\eqref{eq:timeEvo}, we utilize a mean-field approach similar to that formulated in Refs.~\cite{horowicz2021critical,happer2010optically} to simulate the spin dynamics of the $16\times 16$ density matrix $\rho$ for a single atom. The density matrix
\begin{equation}
\rho=\begin{pmatrix}
\rho_{gg} & \rho_{ge}\\
\rho_{eg} & \rho_{ee}
\end{pmatrix}
\end{equation}
is defined in terms of the eight spin states in the $5^2\text{S}_{1/2}$ $F_g=1,2$ hyperfine manifolds, denoted $\rho_{gg}$, and the eight spin states in the $5^2\text{P}_{1/2}$ $F_e=1,2$ hyperfine manifolds, denoted $\rho_{ee}$. The optical coherences are described by $\rho_{eg}$ and $\rho_{ge}$ submatrices.

\begin{figure*}[!tbh]\centering
\includegraphics[width=.83\textwidth]{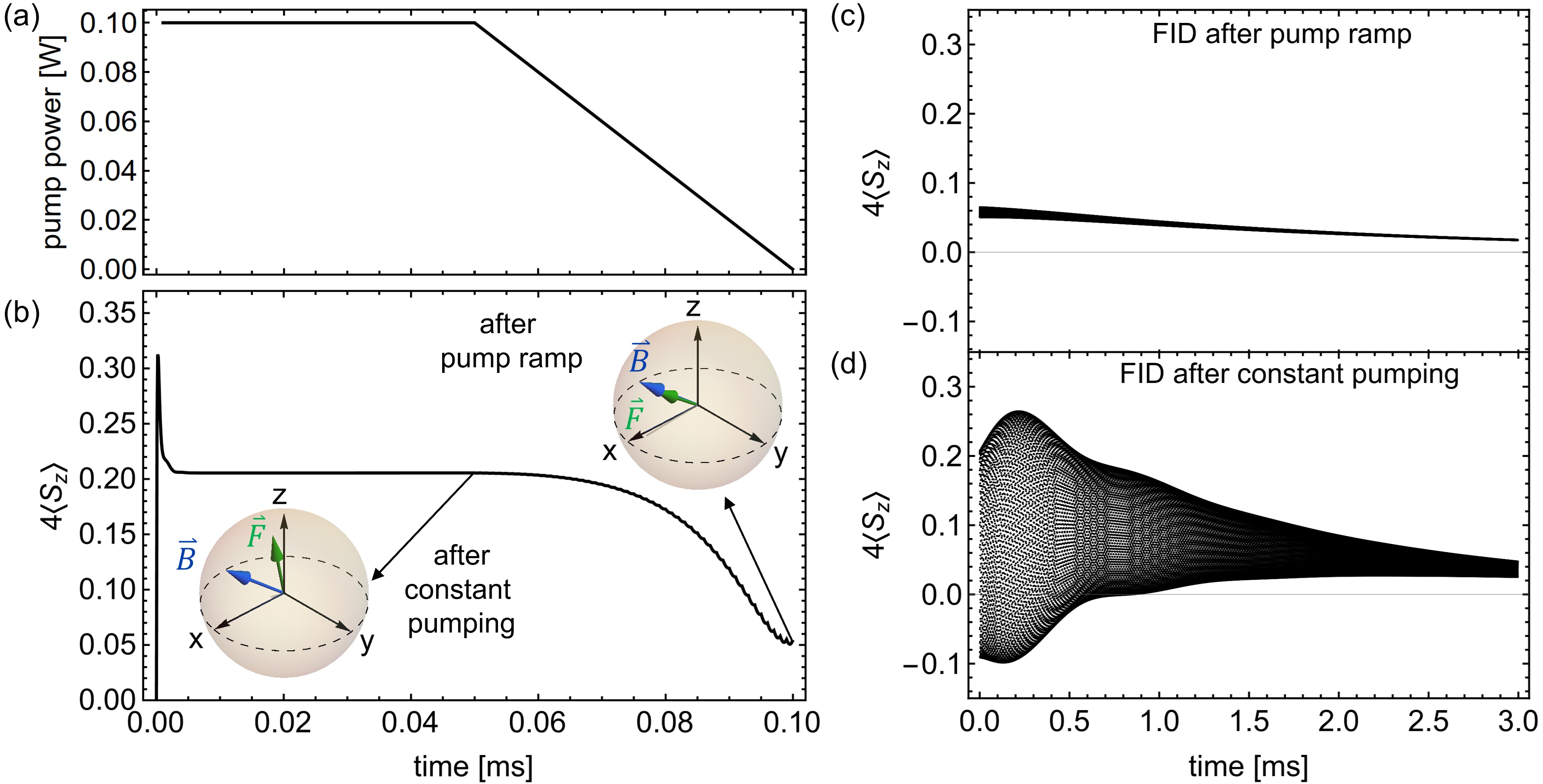}
\caption{Adiabatic optical pumping simulation for $(\alpha,\beta)=(0^{\circ},51.6^{\circ})$. (a) Optical pump power during the adiabatic optical pumping. (b) Simulated
Faraday rotation angle  $\theta_F \propto 4\langle S_z \rangle$ during optical pumping. The two spheres illustrate the relative orientation between the total atomic spin ($\vec{F}$) and the magnetic field ($\vec{B}$) after pumping at constant optical power and after the pump power ramp. (c) The simulated FID of the atomic state after the pump power ramp, which exhibits a small amount of spin precession due to residual misalignment between $\vec{F}$ and $\vec{B}$. (d) In contrast to (c), the simulated FID after pumping at constant optical power produces a larger spin precession signal.}
\label{fig:pumpSim}
\end{figure*}

Under the RWA with respect to the optical field, the master equation that governs the time evolution of $\rho$ is given by
\begin{equation}
\label{eq:pumpModel}
\frac{\partial \rho}{\partial t}=-\frac{i}{\hbar}[\mathcal{H},\rho]+\mathcal{L}(\rho).
\end{equation}
The Hamiltonian
\begin{equation}
\mathcal{H}=\begin{pmatrix}
\mathcal{H}_{gg} & \mathcal{H}_{ge}\\
\mathcal{H}_{eg} & \mathcal{H}_{ee}
\end{pmatrix}
\end{equation}
 is written in terms of the 5$^2$S$_{1/2}$ Hamiltonian $\mathcal{H}_{gg}$, the 5$^2$P$_{1/2}$ Hamiltonian $\mathcal{H}_{ee}$, and the optical coupling interaction $\mathcal{H}_{ge}=\mathcal{H}_{eg}^{\dagger}$. The 5$^2$S$_{1/2}$ and 5$^2$P$_{1/2}$ Hamiltonians
\begin{align}
\begin{split}
\label{eq:groundStateHam}
\mathcal{H}_{gg}=&A_g\mathbf{I}_g\cdot\mathbf{S}_g+\mu_{B}(g_s^{(g)}\mathbf{S}_g+g_i^{(g)}\mathbf{I}_g)\cdot\vec{B}
\\&-\mathcal{I}_g \mathcal{E}_{g,\ket{1,0}},
\end{split}
\\
\begin{split}
\label{eq:excitedStateHam}
\mathcal{H}_{ee}=&A_e\mathbf{I}_e\cdot\mathbf{S}_e+\mu_{B}(g_s^{(e)}\mathbf{S}_e+g_i^{(e)}\mathbf{I}_e)\cdot\vec{B}
\\&-\mathcal{I}_e (\mathcal{E}_{e,\ket{1,0}}+h\delta),
\end{split}
\end{align}
consist of the hyperfine interaction $A\mathbf{I}\cdot\mathbf{S}$ and Zeeman interaction $\mu_{B}(g_s\mathbf{S}+g_i\mathbf{I})\cdot\vec{B}$. The hyperfine coupling constants and Land\'e $g$ factors are given by $A_g=3.417$ GHz and $g^{(g)}_s=2.00232$ for $\mathcal{H}_{gg}$ and $A_e=0.4083$ GHz and $g^{(e)}_s=0.6659$ for $\mathcal{H}_{ee}$. In both cases $g^{(g)}_i=g^{(e)}_i=-0.00099514$. The total electron and nuclear-spin operators for the ground manifolds are denoted $\mathbf{S}_g$ and $\mathbf{I}_g$, respectively, with similar definitions for the excited-state spin operators. In Eqs.~\eqref{eq:groundStateHam} and \eqref{eq:excitedStateHam}, $\mathcal{I}_g$ ($\mathcal{I}_e$) is the identity operator for the $\ket{F_g,m_{F_g}}$ ($\ket{F_e,m_{F_e}}$) basis, and $\delta$ represents the optical detuning from the $\ket{F_g=1,0}\leftrightarrow \ket{F_e=1,0}$ transition. To enforce this resonance condition, we adjust the overall energies of $\mathcal{H}_{gg}$ and $\mathcal{H}_{ee}$ such that, when the optical detuning is on resonance ($\delta=0$), the energy of the $\ket{F_g=1,0}$ state ($\mathcal{E}_{g,\ket{1,0}}$) is equal to the energy of the $\ket{F_e=1,0}$ state ($\mathcal{E}_{e,\ket{1,0}}$).

The optical coupling $\mathcal{H}_{ge}=\mathcal{H}_{eg}^{\dagger}$ is defined in terms of a complex electric field
\begin{equation}
\label{eq:complexElectricField}
\vec{E}=\{E_x,E_ye^{-i\psi},0\}
\end{equation}
of the pump laser and the electric dipole transition operator
\begin{align}
\begin{split}
\bra{F_g,m_{F_g}}\mathcal{H}_{ge}&\ket{F_e,m_{F_e}}=\\&\frac{E_k}{2}\bra{F_g,m_{F_g}}er_k\ket{F_e,m_{F_e}}.
\end{split}
\end{align}
Here $k=\sigma^{\pm},\pi$ denotes transitions characterized by $m_{F_e}=m_{F_g}\pm1$ for $k=\sigma^{\pm}$ and $m_{F_e}=m_{F_g}$ for $k=\pi$. It is convenient to write $\vec{E}$ in a spherical basis, namely $E_{\pm}=\vec{E}\cdot \epsilon_{\mp}$ and $E_{\pi}=\vec{E}\cdot \epsilon_{\pi}$ with $\epsilon_\pm=\{ \frac{1}{\sqrt{2}},\pm \frac{i}{\sqrt{2}},0\}$ and $\epsilon_\pi=\{0,0,1\}$. To simulate realistic optical pumping, we use pump polarization parameters  $E_y/E_x=1.68$ and $\psi=-0.8$ rad measured of our pump beam in front of the vapor cell using a polarization analyzer. The electric field norm is estimated from the laser power $P$ through
\begin{equation}
|\vec{E}|=\sqrt{\frac{4P}{\pi w^2 c \epsilon_0}}
\end{equation}
assuming a Gaussian waist $w=1.5$ mm.

The relaxation superoperator $\mathcal{L}(\rho)$, expressed as 
\begin{equation}
\mathcal{L}(\rho)=\begin{pmatrix}
\mathcal{L}_{gg}(\rho)  & \mathcal{L}_{ge}(\rho) \\
\mathcal{L}_{eg}(\rho)  & \mathcal{L}_{ee}(\rho)
\end{pmatrix},
\end{equation}
describes relaxation through radiation and collision channels in the ground and excited states, as well as optical broadening $\mathcal{L}_{eg}(\rho)=\mathcal{L}_{ge}(\rho)^{\dagger}=-\rho_{eg}\Gamma_o/2$ with linewidth $\Gamma_o/2\pi=5.6$ GHz due to Rb-N$_2$ collisions. 

Excited-state relaxation is expressed as
\begin{equation}
\label{eq:excitedRelax}
\mathcal{L}_{ee}(\rho)=-(\Gamma_s+\Gamma_q)\rho_{ee}-\Gamma_p(\rho_{ee}-\phi_{ee})
\end{equation}
where $\Gamma_p$ is the excited-state spin-destruction (SD) rate from Rb-N$_2$ collisions and $\Gamma_q$ is the de-excitation rate (quenching) from Rb-N$_2$ collisions. For the high quenching rates at our vapor cell parameters, listed in Table~\ref{tab:collisionRatesHeading}, de-excitation due to spontaneous emission at the rate $\Gamma_s=36.1\times 10^7$ s$^{-1}$ is negligible. However, we include it in Eq.~\eqref{eq:excitedRelax} for completeness. The term $\phi_{ee}$ ($\phi_{gg}$) is defined in a manner akin to $\phi$ as presented in Eq.~\eqref{eq:timeEvo}, with the distinction that it is expressed using $\rho_{ee}$ ($\rho_{gg}$).

Ground manifold relaxation and repopulation is given by~\cite{jau2005new}
\begin{align}
\begin{split}
\label{eq:groundRelaxation}
\mathcal{L}_{gg}(\rho)=&-\Gamma_{se}(\rho_{gg}-\phi_{gg}(1+4\langle \mathbf{S}_g\rangle\cdot \mathbf{S}_g))
\\&-\Gamma_{sd}(\rho_{gg}-\phi_{gg})-\Gamma_{\text{D}}(\rho_{gg}-\rho_{gg}^{e})\\&+\frac{4}{3}\Gamma_s (\phi_{ee}-\frac{\rho_{ee}}{4})+\Gamma_q\phi_{ee}
\end{split}
\end{align}
\\where
$\Gamma_{\text{sd}}$ is the SD rate due to Rb-N$_2$ and Rb-Rb collisions, $\Gamma_{\text{se}}$ is the SE collision rate, and $\Gamma_{\text{D}}$ is the wall collision rate tabulated in Table~\ref{tab:collisionRatesHeading}. The term $\rho_{gg}^{e}$ is the same equilibrium density matrix used in Eq.~\eqref{eq:timeEvo}. The last two terms of Eq.~\eqref{eq:groundRelaxation} describe repopulation of the ground manifold due to spontaneous emission and quenching. 

\section{Ramsey and Rabi FS simulation}
\label{sec:RamseyRabiFSSimulation}

To assess the magnitude of potential systematic errors in Ramsey and Rabi FS, we conduct simulations of Ramsey and Rabi Faraday rotation signals, using predetermined values for the magnetic field strength $(B=50$ $\mu$T) and the pressure shift $(\nu_{\text{bg}}=88$ kHz), alongside MPE parameters calibrated from Rabi measurements [see Table~\ref{tab:PEfreqDep}].

\begin{figure}[!tbh]\centering
\includegraphics[width=.35\textwidth]{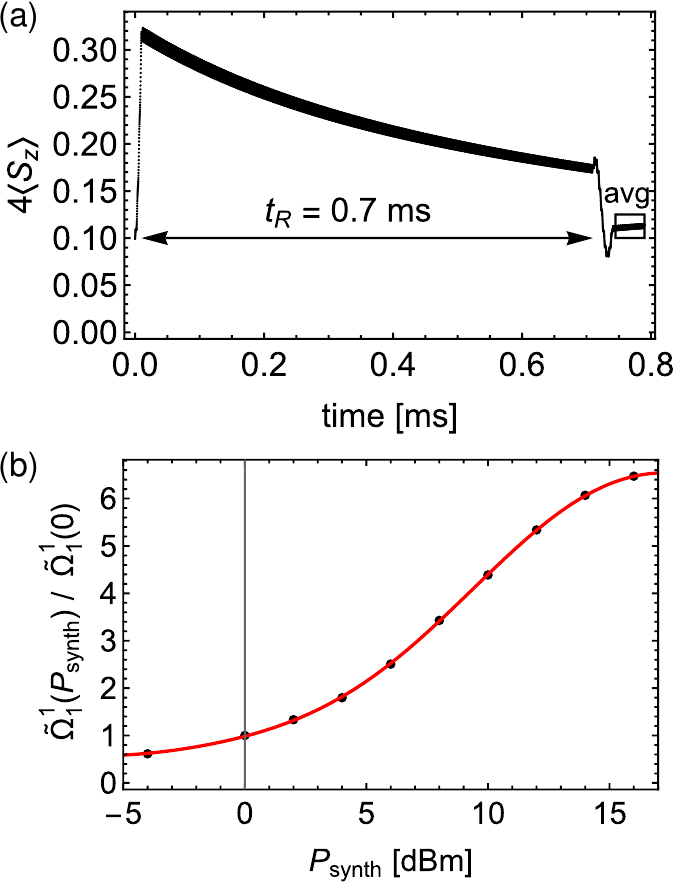}
\caption{(a) Simulated Faraday rotation angle ($\theta_F \propto 4 \langle S_z \rangle$) during the Ramsey pulse sequence for $t_R=0.7$ ms at the magnetic field direction $(\alpha,\beta)=(0,34^{\circ})$. (b) The relative change of the $\pi$ generalized Rabi frequency ($\tilde{\Omega}_{m=1}^{m^{\prime}=1}$), with $\Delta_{m=1}^{m^{\prime}=1}\approx 0$, at different microwave power settings ($P_{\text{synth}}$). The red line is generated from polynomial interpolation, and represents the relative change of the microwave field amplitude ($|\vec{\mathcal{B}}|$) for different power settings of the microwave synthesizer.}
\label{fig:ramseySimWithWindFreak}
\end{figure}

\begin{figure*}[!tbh]\centering
\includegraphics[width=.99\textwidth]{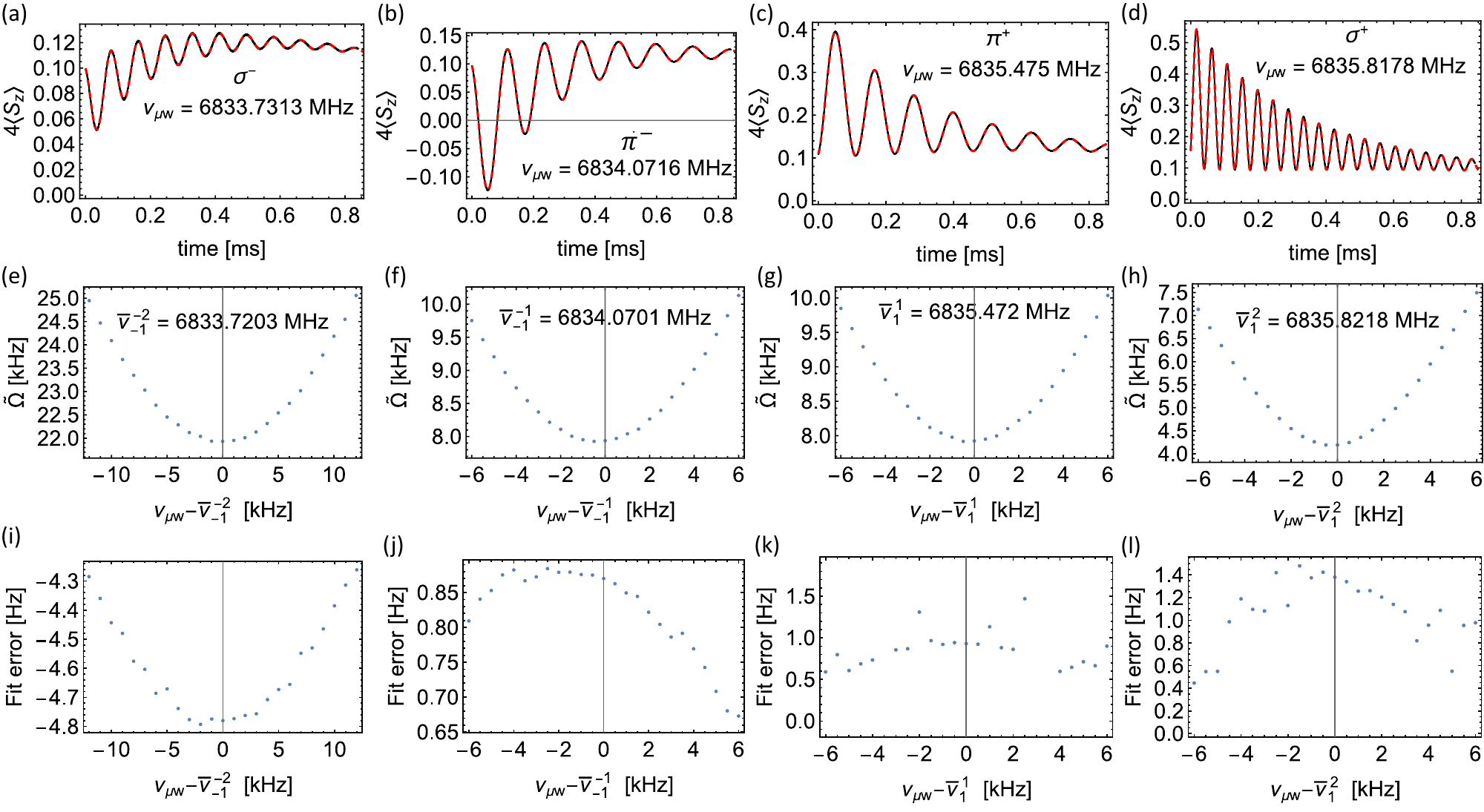}
\caption{(a-d) Simulated Rabi oscillations (black) and fits (red dashed) for $(\alpha,\beta)=(0,34^{\circ})$. (e-h) Detuning dependence of the generalized Rabi frequency fits about $\overline{\nu}_m^{m^{\prime}}$. (i-l) Time-domain frequency fit errors evaluated by comparing the $\delta \lambda_m^{m^{\prime}}$ model to the fitted generalized Rabi frequencies. }
\label{fig:theoreticalRabFits}
\end{figure*}

\begin{figure}[!tbh]\centering
\includegraphics[width=.48\textwidth]{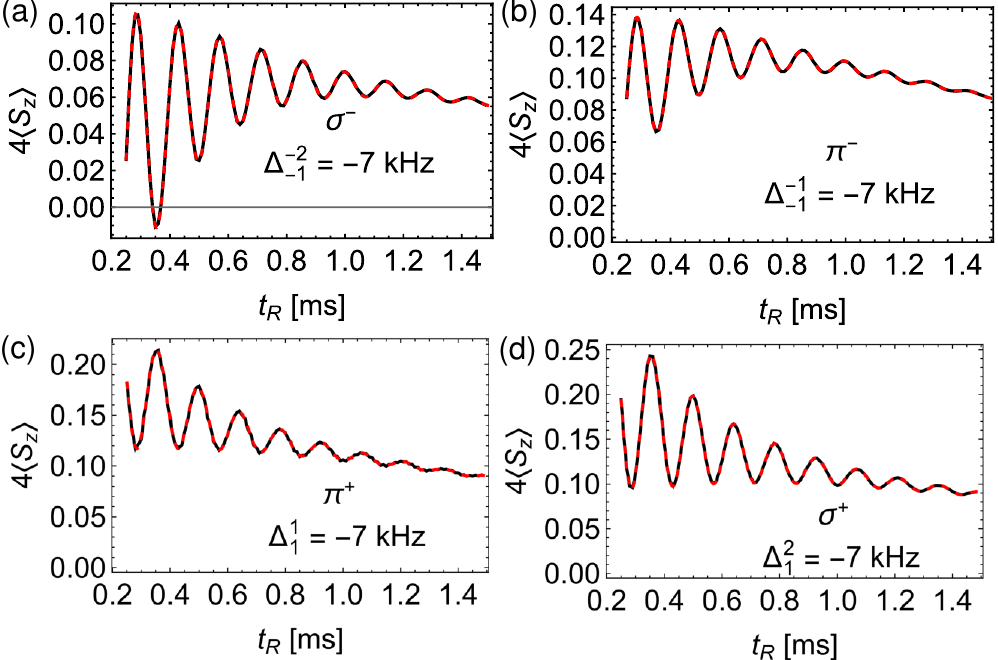}
\caption{Simulated Ramsey fringes (black) for $(\alpha,\beta)=(0,34^{\circ})$. Time-domain fits (red dashed), using Eq.~\eqref{eq:timeDomainFit}, extract the Ramsey fringe frequency.}
\label{fig:ramseyFringeSim}
\end{figure}

First, $D_1$ optical pumping is simulated using experimental pump beam parameters to estimate the initial atomic density matrix $\rho_0$, which describes the ground state 5$^2$S$_{1/2}$ manifold. This simulation, using the optical pumping model discussed in Appendix~\ref{sec:fullopticalPumpModel}, is conducted across the same magnetic field directions as those probed in Fig.~\ref{fig:finalHeadingErr}. With this model, we verify the spin-aligning effect of AOP by comparing simulated FID signals initialized with either constant optical power pumping or linearly ramped-off optical power for a $50$-$\mu$T magnetic field in the direction $(\alpha,\beta) = (0^{\circ}, 51.6^{\circ})$ [see Fig.~\ref{fig:pumpSim}]. For this magnetic field direction, the total atomic spin vector $\vec{F}=\langle \mathbf{F} \rangle$ after $50$ $\mu$s of constant pumping has a steady-state direction given by $(-15^{\circ},11^{\circ})$. Meanwhile the steady-state spin vector direction after also applying the $50$ $\mu$s linear power ramp, as employed during AOP, is $(4^{\circ},53^{\circ})$, which closely aligns with the magnetic field direction.

\begin{figure}[!tbh]\centering
\includegraphics[width=.42\textwidth]{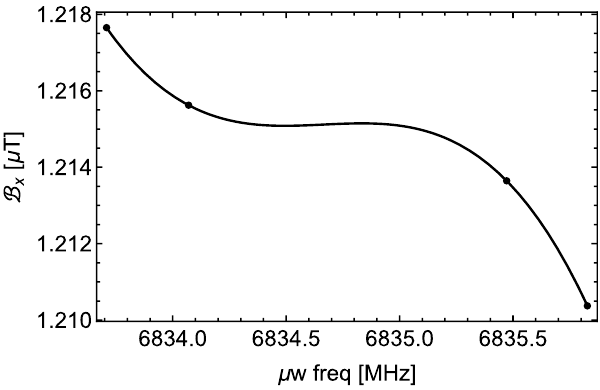}
\caption{Estimated microwave frequency dependence of the MPE-1 $\mathcal{B}_x$ component. The frequency dependence is obtained by performing polynomial interpolation (black line) on the calibrated MPE-1 $\mathcal{B}_x$ parameters evaluated at four different microwave frequencies listed in Table~\ref{tab:PEfreqDep} (black dots). The frequency dependence of the other microwave parameters are estimated similarly.}
\label{fig:fakeFreqDep}
\end{figure}

\begin{figure*}[!tbh]
\begin{center}
\includegraphics[width=.9\textwidth]
{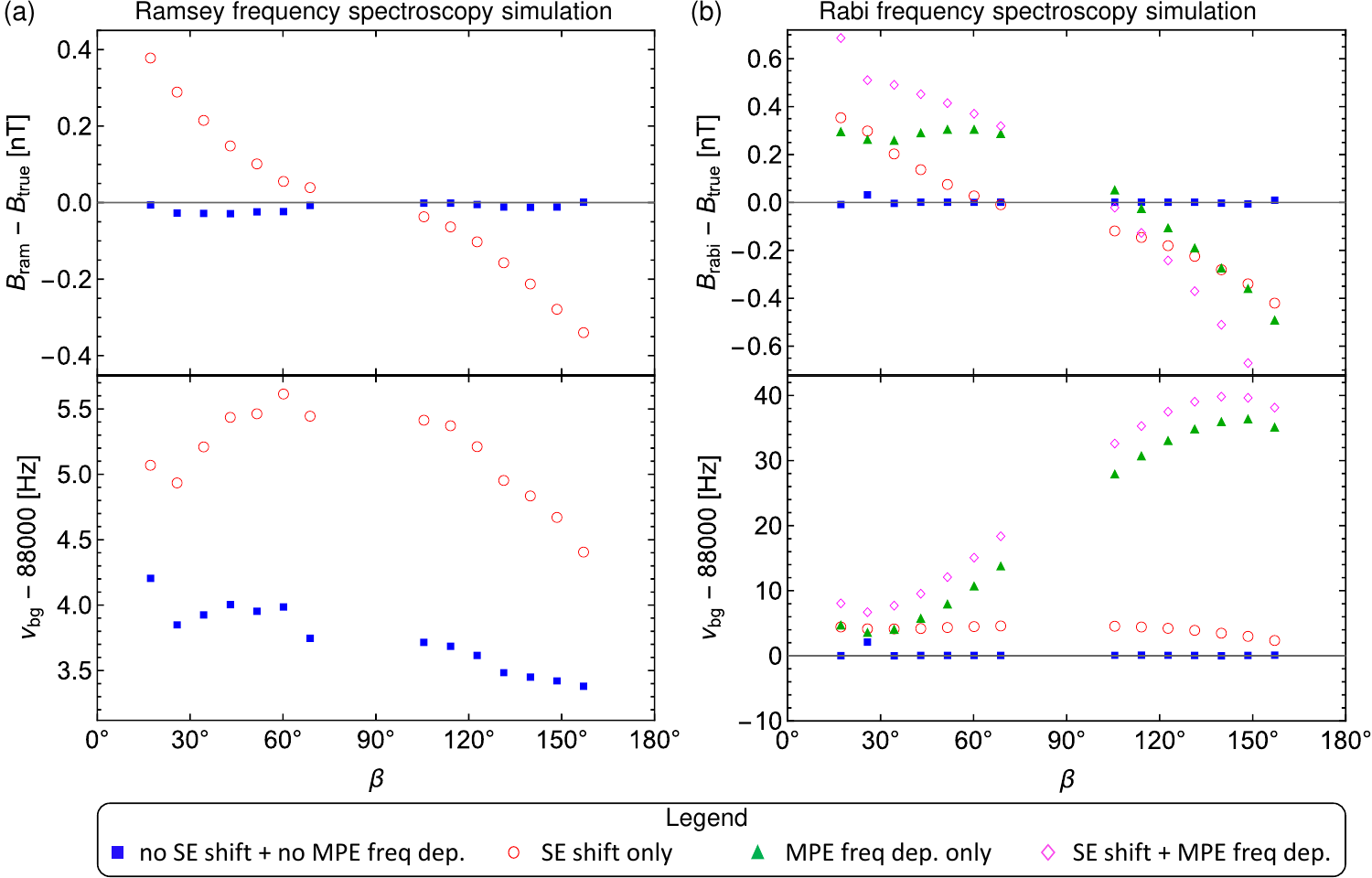}
\end{center}
\caption{Simulated magnetic field strength and pressure shift errors using Ramsey and Rabi frequency spectroscopy (FS). (a) Errors in Ramsey FS. If no spin-exchange (SE) frequency shifts, errors from off-resonant microwave driving are contained within 50 pT (blue). SE frequency shifts at our vapor cell temperature produce errors within 400 pT across different dc field directions. (b)  Errors in Rabi FS. Shown are the average magnetic field strengths deduced from Rabi measurements across all four hyperfine transitions and all three MPEs. With no MPE frequency dependence and SE frequency shifts, the magnetic field strength errors are within 50 pT (blue). Errors from SE frequency shifts are at a similar level as the Ramsey measurements. The MPE frequency dependence [see Fig.~\ref{fig:fakeFreqDep}] contributed a major source of systematics at the 0.4 nT level (green).}
\label{fig:simTotalHeadingError}
\end{figure*}

For the Ramsey FS simulations, we use Eq.~\eqref{eq:timeEvo} to simulate the $\pi/2-t_R-3\pi/2$ Ramsey sequence, as shown in Fig.~\ref{fig:ramseySimWithWindFreak}(a). The microwave field is defined using the MPE-1 microwave parameters, which correspond to the same cavity mode excitations used in the Ramsey measurements. In the simulations, the microwave parameters are taken to be the average of the values listed in the first four rows of Table~\ref{tab:PEfreqDep}. In the experiment, the power of the microwave synthesizer ($P_{\text{synth}}$) at each hyperfine transition was manually optimized to satisfy the 10-$\mu$s $\pi /2$ time in the Ramsey measurements. To mimic this process in the Ramsey simulations, we assume that changing $P_{\text{synth}}$ does not affect the MPE structure, but only affects the microwave amplitude given by $|\vec{\mathcal{B}}|=\sqrt{\mathcal{B}_x^2+\mathcal{B}_y^2+\mathcal{B}_z^2}$. To characterize the dependence of $|\vec{\mathcal{B}}|$ on $P_{\text{synth}}$, we measure the generalized Rabi frequency at $\nu_{\mu \text{w}}=\overline{\nu}_{m=1}^{m^{\prime}=1}$ ($\tilde{\Omega}_{m=1}^{m^{\prime}=1}$) as a function of $P_{\text{synth}}$ [see Fig.~\ref{fig:ramseySimWithWindFreak}(b)]. Here we assume that $\tilde{\Omega}_{m=1}^{m^{\prime}=1}$ is proportional to $|\vec{\mathcal{B}}|$. Thus, from knowledge of the $P_{\text{synth}}$ settings in the Ramsey measurements, it is possible to use comparable MPE parameters in the simulations. As a representative example of the simulated signals, Fig.~\ref{fig:ramseyFringeSim} displays simulated Ramsey fringes at $\beta=34^{\circ}$ with 7-kHz microwave detuning from each hyperfine transition.

For Rabi FS simulations, we similarly use Eq.~\eqref{eq:timeEvo} to generate Rabi oscillations at the same microwave frequencies and MPEs used in the measurements. Realistic frequency dependence of the MPE parameters is estimated using polynomial interpolation [see Fig.~\ref{fig:fakeFreqDep}] of the calibrated parameters listed in Table~\ref{tab:PEfreqDep}. We also assume a drift of 0.4 \% of the microwave field amplitude in these simulations to mimic possible experimental drift. This drift appears randomly in the analysis because the microwave frequencies are taken in random order, and hence causes negligible systematic errors.   

Examples of simulated Rabi oscillations and corresponding generalized Rabi frequency fits at $\beta=34^{\circ}$ are shown in Fig.~\ref{fig:theoreticalRabFits}(a-h). We observe small systematic errors in these fits of the order of a few hertz as shown in Fig.~\ref{fig:theoreticalRabFits}(i-l). These errors likely arise from the nontrivial lineshape deviations of the simulated Rabi oscillations from the exponential-decay fitting model (Eq.~\eqref{eq:timeDomainFit}) due to atomic collisions. As shown in Fig.~\ref{fig:simTotalHeadingError}, however, these fitting errors only cause scalar systematics within 50 pT. 

Systematic errors [see Fig.~\ref{fig:simTotalHeadingError}] for Ramsey and Rabi FS are estimated by comparing the fitted magnetic field strengths and pressure shifts of this simulated data to the known magnetic field strength $B=50$ $\mu$T and pressure shift $\nu_{\text{bg}}=88$ kHz. Besides systematic shifts caused by off-resonant driving, Fig.~\ref{fig:simTotalHeadingError} also reveals the impact of microwave frequency dependence and SE frequency shifts, effects that we can isolate in our modeling.

Our simulations indicate that Ramsey FS off-resonant driving and time-domain fitting errors lead to systematic errors within $50$ pT; see the blue squares in Fig.~\ref{fig:simTotalHeadingError}. Rabi FS shows similar systematic errors, which likely arise from the time-domain fitting errors attributed to collisional lineshape distortions shown in Fig.~\ref{fig:theoreticalRabFits}. Including spin-exchange frequency shifts (red circles in Fig.~\ref{fig:simTotalHeadingError}) increases the Ramsey and Rabi systematic shifts to similar levels within $\pm$ 0.4 nT. 

For Rabi FS, simulated frequency dependence of the MPE parameters causes a systematic error ($\pm0.4$ nT) that is similar in magnitude to the SE frequency shifts (green triangles in Fig.~\ref{fig:simTotalHeadingError}). Microwave frequency dependence could be mitigated with improved flatness of the microwave-cavity mode, or could be compensated by performing MPE calibrations at each microwave frequency used in Rabi FS. MPE frequency dependence is not expected to be a systematic for the Ramsey protocol because Ramsey fringes are fit at a single microwave frequency.

As displayed in Fig.~\ref{fig:simTotalHeadingError}, pressure shifts are more accurately measured with Rabi FS than Ramsey FS when microwave frequency dependence of MPE parameters and spin-exchange frequency shifts are not included. This results from imperfect compensation of frequency shifts from off-resonant driving in Ramsey FS. However, from the MPE frequency dependence assumed in our simulations, Rabi FS pressure shift errors can reach up to $40$ Hz. From this result, we expect that the 10-Hz scale discrepancies in Fig.~\ref{fig:rabMagDifferentTrans} between pressure shift measurements of different MPEs are likely because of MPE frequency dependence.

\begin{figure}[!tbh]\centering
\includegraphics[width=.43\textwidth]{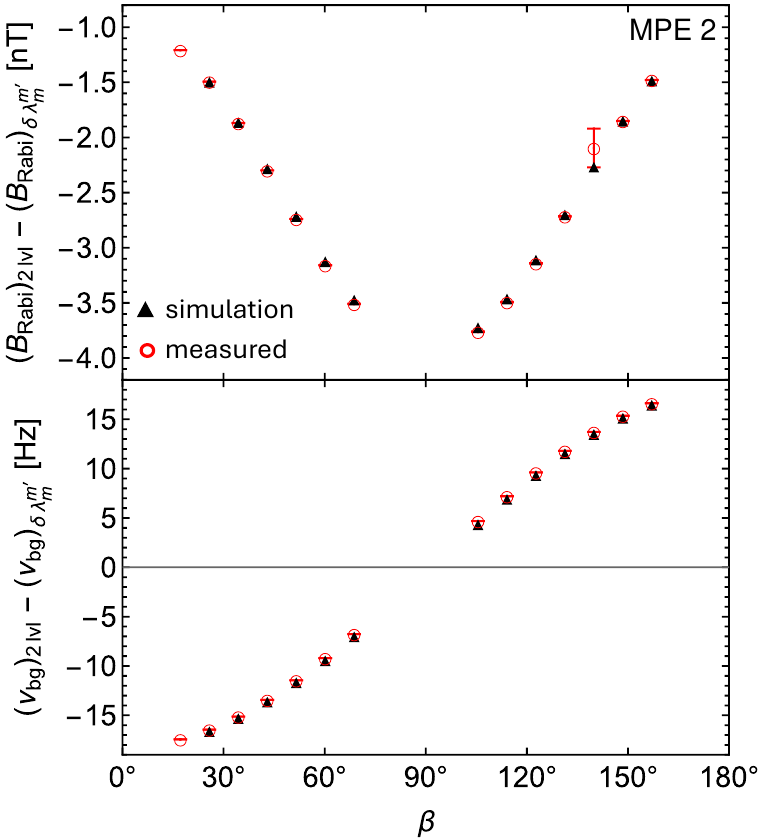}
\caption{Difference between the magnetic field strength $B$ and pressure shift $\nu_{\text{bg}}$ fits, using the two-level Rabi model, given by Eq.~\eqref{eq:2lvl}, against the full Hamiltonian model, given by Eq.~\eqref{eq:rabiLambda}, evaluated on simulated (black triangle) and measured (red circle) data. As a representative example, these fits only utilize the $\sigma^{\pm}$ hyperfine transitions with the MPE-2 Rabi data. This comparison confirms that failing to accurately model off-resonant driving would have led to errors in magnetic field strength measurements up to 4 nT.}
\label{fig:compareWithoutHam}
\end{figure}

In Fig.~\ref{fig:compareWithoutHam}, we show Rabi FS systematic errors that would arise if the atom-microwave coupling of off-resonant transitions were not properly modeled. This is done by comparing fits of the magnetic field strength $B$ and the pressure shift $\nu_{\text{bg}}$ using the two-level Rabi formula, defined in Eq.~\eqref{eq:2lvl}, against using the $\delta \lambda_m^{m^{\prime}}$ model, defined in Eq.~\eqref{eq:rabiLambda}. We denote the magnetic field strengths obtained from these two different fitting models as $(B_{\text{Rabi}})_{2\text{lvl}}$ and $(B_{\text{Rabi}})_{\delta \lambda^{m^{\prime}}_m}$ with similar notation for fits of $\nu_{\text{bg}}$. In Fig.~\ref{fig:compareWithoutHam}, we observe nearly perfect agreement of the discrepancy between the two-level model and the $\delta \lambda ^{m^{\prime}}_m$ model by comparing the difference $(B_{\text{Rabi}})_{2\text{lvl}}-(B_{\text{Rabi}})_{\delta \lambda^{m^{\prime}}_m}$ using simulated and measured data. This comparison leads us to conclude that inadequate modeling of off-resonant driving, i.e., not using the $\delta \lambda_m^{m^{\prime}}$ model, would result in magnetic field strength errors up to 4 nT.

\section{FID heading error simulation}
\label{sec:FIDHeadingErrorSim}
\begin{figure}[!tbh]
\begin{center}
\includegraphics[width=.43\textwidth]
{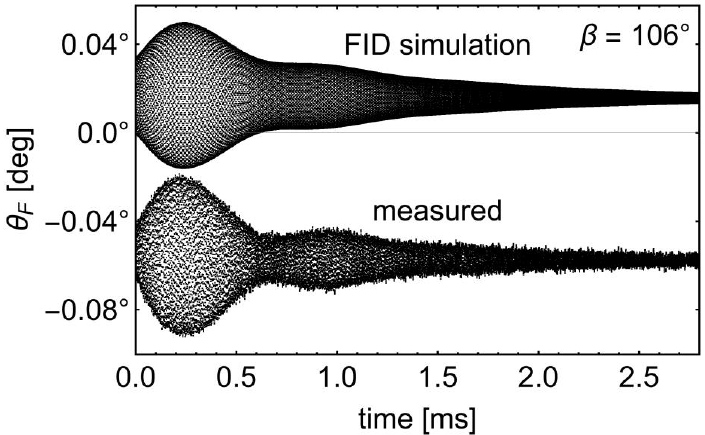}
\end{center}
\caption{Measured FID signal, averaged over three traces with a moving average of ten data points, at $\beta=106^{\circ}$, overlaid with the corresponding scaled simulated FID for comparison.}
\label{fig:simFIDCompare}
\end{figure}

To simulate FID heading error we use the optical pumping model described in Appendix~\ref{sec:fullopticalPumpModel} to estimate the ground-state $8\times 8$ density matrix $\rho_0$ at the start of FID measurements. In these simulations we assume the same experimental pump settings of 100-$\mu$s pulse duration and an optical power of $P=0.4$ W. After pumping, we free evolve the atomic state for 3 ms to generate an FID Faraday rotation signal using Eq.~\eqref{eq:timeEvo} and Eq.~\eqref{eq:faradayRotationOperator}. Figure~\ref{fig:simFIDCompare} displays an FID measurement taken at the magnetic field direction $(\alpha,\beta) = (0^\circ, 106^\circ)$, overlaid with the corresponding simulated FID signal, and scaled to the same size for comparison.

\begin{figure}[!tbh]
\begin{center}
\includegraphics[width=.43\textwidth]
{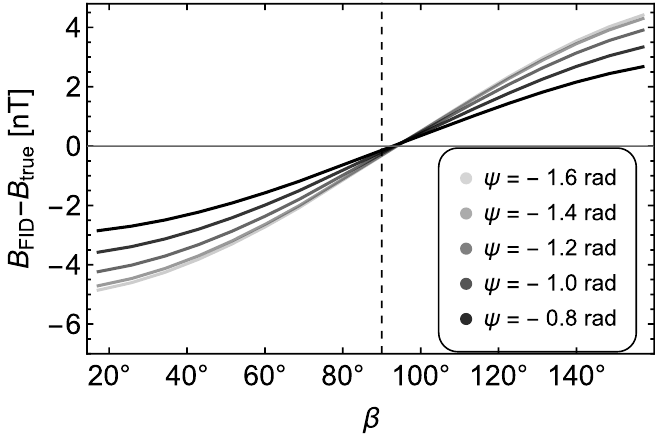}
\end{center}
\caption{Simulated FID heading error for different pump relative phases $\psi$ between $E_x$ and $E_y$ electric field components defined in Eq.~\eqref{eq:complexElectricField}.}
\label{fig:headErrSimFID}
\end{figure}

To evaluate the simulated heading error, plotted in Fig.~\ref{fig:headErrSimFID}, we fit the simulated FID signals with the same fitting protocol described in Eq.~\eqref{eq:FSP}. The discrepancy in the size of the beating observed in the simulated FID signal in Fig.~\ref{fig:simFIDCompare}, which arises from the different spin precession frequencies between the $F=1$ and $F=2$ manifolds, compared to the beating in the FID measurement indicates that the heading error simulations do not perfectly model the experiment at the estimated parameters. This mismatch could arise from uncertainty in the pump beam parameters. For example, there is some uncertainty in the pump electric field polarization at the location of the atoms due to optical reflections off the uncoated glass walls of the vapor cell that create a small etalon. In addition, there is some uncertainty in the pump optical frequency of the order of a few gigahertz due to wavemeter uncertainty and drift. Furthermore, spatial dependence of the pump power inside the vapor cell due to absorption is not considered here. We vary the pump relative phase from $\psi=-0.8$ rad, defined in Eq.~\eqref{eq:complexElectricField}, within the FID heading error simulations to show, as one example, the effect of these pump beam parameters. Despite some ambiguity in the pump beam parameters, our simulations in Fig.~\ref{fig:headErrSimFID} produce qualitatively similar heading errors to those predicted from Rabi and Ramsey FS measurements in Fig.~\ref{fig:finalHeadingErr}, except for a difference in the overall offset mentioned in the main text.

\bibliography{refs.bib}
\end{document}